\RequirePackage{caption}
\let\abovecaptionskip\relax
\let\belowcaptionskip\relax
\documentclass[runningheads]{llncs}
\usepackage[usenames,dvipsnames]{xcolor}
\usepackage{graphicx}

\usepackage{tikz}
\usepackage{comment}
\usepackage{amsmath,amssymb} 
\usepackage{color}

\graphicspath{{figures/}}

\DeclareMathOperator*{\argmax}{arg\,max}

\DeclareMathOperator*{\sign}{sgn}
\DeclareMathOperator*{\proj}{proj}
\usepackage{hyperref}
\usepackage{cleveref}
\crefname{figure}{Fig.}{Figs.}
\crefname{equation}{Eq.}{Eqs.}
\usepackage{bm}
\usepackage{siunitx}
\usepackage{wrapfig}
\usepackage{adjustbox}
\usepackage{floatrow}
\usepackage{subfig}
\usepackage{pdflscape}
\usepackage[para,online]{threeparttable}
\usepackage{array, makecell}
\usepackage{booktabs} 
\usepackage{multirow}
\usepackage{microtype}

\tikzset{
vertex/.style = {
circle,
fill = black,
outer sep = 2pt,
inner sep = 1pt, } }
\usetikzlibrary{calc,arrows.meta,decorations.pathmorphing,backgrounds,fit,positioning,shapes.symbols,chains,shapes,matrix}
\usetikzlibrary{shapes,shapes.multipart,decorations.pathreplacing}
\usetikzlibrary{calc}

\tikzset{trapezium stretches=true}
\tikzset{
    ncbar angle/.initial=90,
    ncbar/.style={
        to path={
           (\tikztostart)
        -- ($(\tikztostart)!#1!\pgfkeysvalueof{/tikz/ncbar angle}:(\tikztotarget)$)
        -- ($(\tikztotarget)!($(\tikztostart)!#1!\pgfkeysvalueof{/tikz/ncbar angle}:(\tikztotarget)$)!\pgfkeysvalueof{/tikz/ncbar angle}:(\tikztostart)$)
        -- (\tikztotarget)
    }},
    ncbar/.default=.5cm
}

\hypersetup{
colorlinks=true,
linkcolor=blue,
citecolor=blue,
filecolor=blue,
urlcolor=blue}

\begin{document}

\pagestyle{headings}
\mainmatter
\def\ECCVSubNumber{48}  

\title{Conditional Adversarial Camera Model Anonymization} 

\titlerunning{Conditional Adversarial Camera Model Anonymization}
%
\author{Jerone T. A. Andrews \and
Yidan Zhang \and
Lewis D. Griffin}
\authorrunning{J. T. A. Andrews et al.}
%
\institute{Department of Computer Science, University College London
\email{jerone.andrews@cs.ucl.ac.uk}}
\maketitle

\begin{abstract}
The model of camera that was used to capture a particular photographic image (model attribution) is typically inferred from high-frequency model-specific artifacts present within the image. Model anonymization is the process of transforming these artifacts such that the apparent capture model is changed. We propose a conditional adversarial approach for learning such transformations. In contrast to previous works, we cast model anonymization as the process of transforming both high and low spatial frequency information. We augment the objective with the loss from a pre-trained dual-stream model attribution classifier, which constrains the generative network to transform the full range of artifacts. Quantitative comparisons demonstrate the efficacy of our framework in a restrictive non-interactive black-box setting.

\keywords{Camera model anonymization, conditional generative adversarial nets, adversarial training, non-interactive black-box attacks, image editing/manipulation, camera model attribution/identification}
\end{abstract}

\section{Introduction}
\label{sec:intro}

Photographic images can be attributed to the specific camera model used for capture~\cite{geradts2001methods}. Attribution is facilitated by inferring model-specific digital acquisition and processing artifacts present within high-frequency pixel patterns~\cite{tuama2015source,marra2015evaluation,filler2008using,gloe2012feature,tuama2016camera,bayar2017design}. While such artifacts have been used to verify the origin and integrity of images, attribution evidently raises concerns about unjustifiable misuse. This is particularly pertinent to individuals such as human rights' activists, photojournalists and whistle-blowers, that reserve the right to privacy and anonymity~\cite{dirik2014analysis}.

In this work, we are not concerned with attribution per se, but the challenging problem of camera model anonymization~\cite{chen2018mislgan,mandelli2017inpainting,karakuccuk2015adaptive}. Model anonymization is the process of transforming model-specific artifacts s.t. the apparent capture model is changed. Namely, the goal is to learn a function that transforms the innate model-specific artifacts of an image to those of a disparate target model. Such a system could then be used to preserve privacy, or conversely for validating the robustness and reliability of attribution methods, particularly when attribution results are admitted as forensic evidence in civil or criminal cases.

\begin{figure}[t]
\centering
\begin{tikzpicture}
[font=\scriptsize\fontfamily{phv}\selectfont]
\node [rectangle, minimum width=30pt, text badly centered, minimum height=30pt, text width=60pt, draw, rounded corners=2pt, very thick, inner sep=0pt, outer sep=0pt, fill=SeaGreen!5] (Lens) at (30pt,125pt) {Lens and Optical Filter(s)};

\node [rectangle, minimum width=30pt, text badly centered, minimum height=30pt, text width=60pt, draw, rounded corners=2pt, very thick, inner sep=0pt, outer sep=0pt,fill=SeaGreen!20] (CFA) at (100pt,125pt) {Color Filter Array (CFA)};

\node [rectangle, minimum width=30pt, text badly centered, minimum height=30pt, text width=60pt, draw, rounded corners=2pt, very thick, inner sep=0pt, outer sep=0pt,fill=SeaGreen!40] (Sensor) at (170pt,125pt) {Imaging Sensor};

\node [rectangle, minimum width=30pt, text badly centered, minimum height=30pt, text width=60pt, draw, rounded corners=2pt, very thick, inner sep=0pt, outer sep=0pt,fill=SeaGreen!60] (Demosaic) at (240pt,125pt) {Color Interpolation};

\node [rectangle, minimum width=30pt, text badly centered, minimum height=30pt, text width=60pt, draw, rounded corners=2pt, very thick, inner sep=0pt, outer sep=0pt,fill=SeaGreen!80] (PP) at (310pt,125pt) {Post-processing};

\draw [-{Triangle[length=4pt,width=4pt]}] (Lens) -- (CFA) [very thick];
\draw [-{Triangle[length=4pt,width=4pt]}] (CFA) -- (Sensor) [very thick];
\draw [-{Triangle[length=4pt,width=4pt]}] (Sensor) -- (Demosaic) [very thick];
\draw [-{Triangle[length=4pt,width=4pt]}] (Demosaic) -- (PP) [very thick];

\node [rectangle, minimum width=30pt, minimum height=50pt, text width=340pt, draw=none, inner sep=0pt, outer sep=0pt,align=justify] (caption) at (170pt,75pt) {Light enters the imaging device via a system of lenses and optical filters. The CFA mosaic projects each pixel to a color: R, G or B. The sensor output is a single-channel mosaic representation. In order to obtain an RGB color image, color interpolation (a.k.a. demosaicing) estimates the missing color information in each channel based on neighboring pixels. Preceding digital storage, the image undergoes various post-processing (e.g. white balancing, color correction, gamma correction, compression).};
\end{tikzpicture}
\caption{Simplified digital image acquisition pipeline}
\label{fig:imageacquisition}
\end{figure}
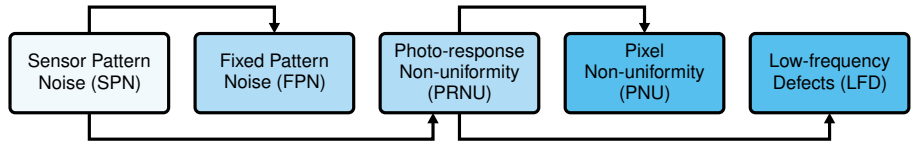

Broadly, previous work on model anonymization tend to view anonymization as \emph{solely} necessitating the attenuation~\cite{mandelli2017inpainting,bonettini2018fooling} or transformation~\cite{chen2018mislgan,chen2019generative,cozzolino2019spoc} of the device-specific pixel non-uniformity imaging sensor noise~\cite{fridrich2009digital}, which is defined as slight variations in the sensitivity of individual pixel sensors. Although initially device-specific (prior to color interpolation), these variations propagate nonlinearly through the processing steps (\cref{fig:imageacquisition}) that result in the final image  and thus end up also depending on model-specific aspects, such as color interpolation, on-sensor signal transfer, sensor design and compression~\cite{fridrich2013sensor}.

Pixel non-uniformity is the dominant noise component of what is termed photo-response non-uniformity (\cref{fig:pattern_noise}). Photo-response non-uniformity noise, however, also contains contributions from low spatial frequency artifacts (independent of the imaging sensor) caused by light refraction on dust particles, optical surfaces and properties of the camera model optics~\cite{lukavs2006detecting}. Such artifacts include optical vignetting~\cite{kirchner2015forensic}, which corresponds to the fall-off in light intensity towards the corners of an image.

\begin{figure}[t]
\centering
\begin{tikzpicture}
[font=\scriptsize\fontfamily{phv}\selectfont]
\node [rectangle, minimum width=30pt, text badly centered, minimum height=30pt, text width=60pt, draw, very thick, inner sep=0pt, outer sep=0pt, fill=Cerulean!5,rounded corners=2pt] (SPN) at (30pt,115pt) {Sensor Pattern Noise (SPN)};

\node [rectangle, minimum width=30pt, text badly centered, minimum height=30pt, text width=60pt, draw, very thick, inner sep=0pt, outer sep=0pt,fill=Cerulean!30,rounded corners=2pt] (FPN) at (100pt,115pt) {Fixed Pattern Noise (FPN)};

\node [rectangle, minimum width=30pt, text badly centered, minimum height=30pt, text width=60pt, draw, very thick, inner sep=0pt, outer sep=0pt,fill=Cerulean!30,rounded corners=2pt] (PRNU) at (170pt,115pt) {Photo-response Non-uniformity (PRNU)};

\node [rectangle, minimum width=30pt, text badly centered, minimum height=30pt, text width=60pt, draw, very thick, inner sep=0pt, outer sep=0pt,fill=Cerulean!60,rounded corners=2pt] (PNU) at (240pt,115pt) {Pixel Non-uniformity (PNU)};

\node [rectangle, minimum width=30pt, text badly centered, minimum height=30pt, text width=60pt, draw, very thick, inner sep=0pt, outer sep=0pt,fill=Cerulean!60,rounded corners=2pt] (LFD) at (310pt,115pt) {Low-frequency Defects (LFD)};

\draw [-{Triangle[length=4pt,width=4pt]}] (30pt,130pt) -- (30pt,140pt) -- (100pt,140pt) -- (100pt,130pt) [very thick];
\draw [-{Triangle[length=4pt,width=4pt]}] (30pt,100pt) -- (30pt,90pt) -- (160pt, 90pt) -- (160pt,100pt)[very thick];

\draw [-{Triangle[length=4pt,width=4pt]}] (170pt,130pt) -- (170pt,140pt) -- (240pt,140pt) -- (240pt,130pt)[very thick];
\draw [-{Triangle[length=4pt,width=4pt]}] (170pt,100pt) -- (170pt,90pt) -- (310pt,90pt) -- (310pt,100pt)[very thick];

\node [rectangle, minimum width=30pt, minimum height=50pt, text width=340pt, draw=none, inner sep=0pt, outer sep=0pt, align=justify] (caption) at (170pt,55pt)
{Imaging SPN is defined as any noise component that survives frame averaging~\cite{holst1998ccd}---i.e. systematic distortions---and primarily comprises of FPN and PRNU. FPN is defined as the pixel-to-pixel differences when the sensor is not exposed to light~\cite{lukavs2006digital}, whereas PRNU is the dominant component of SPN. PRNU is largely caused by PNU, which is defined as stochastic variations in the sensitivity of individual pixel sensors to light. Finally, LFD are artifacts that change slowly in intensity over long spatial distances (e.g. optical vignetting). LFD are a result of the lens and camera optics, as opposed to the imaging sensor.};
\end{tikzpicture}
\caption{Imaging sensor pattern noise sources}
\label{fig:pattern_noise}
\end{figure}

Model anonymization approaches based on pixel non-uniformity invariably suppress the noise-free imaging sensor response (image content), via a denoising filter, and instead work with the \emph{noise residual} (the observed image minus its estimated noise-free image content). This is premised on improving the signal-to-noise ratio between the photo-response non-uniformity noise (signal of interest) and the observable image. However, this precludes the anonymization process from attending to discriminative low spatial frequency model-specific artifacts, since they no longer exist within the high-frequency noise residual.

Despite model anonymization loosely falling under the remit of image editing~\cite{perarnau2016invertible,wang2018face,zhu2016generative}, the targeted anonymizing transformations that we seek should not alter the image content of an input. Minimal distortion is easily achieved by formulating the problem as a lossy reconstruction task, however low distortion is often at odds with high perceptual quality~\cite{blau2018perception}. We therefore absorb the lossy reconstruction task into a simple adversarial training procedure, taking inspiration from conditional generative adversarial networks~\cite{mirza2014conditional}. The gist of the procedure is that the generator transforms (with low distortion) an input image conditioned on a target camera model label, and tries to fool the discriminator into thinking the transformed image's prediction error features (low-level high-frequency pixel value dependency features) are real and coherent with the condition.

In contrast to previous work, we cast model anonymization as the process of transforming both high and low spatial frequency model-specific artifacts. With this in mind, our conditional adversarial camera model anonymizing (Cama) framework also includes a fixed (w.r.t. its parameters) dual-stream discriminative decision-making component (evaluator) that decides whether a transformed image belongs to its target class. Each stream captures a different aspect of the input data. Specifically, we decompose an input into its high and low spatial frequency components and assign each to its own stream. The intuition is that this allows the evaluator to \emph{independently} reason over specific information present in each. Augmenting the adversarial objective with the evaluator's \emph{discriminative} objective reinforces the transformation process and ensures that both high and low spatial frequency model-specific artifacts are attended to by the generator.

Quantitative results underscore the efficacy of our framework when attacking a variety of non-interactive black-box target classifiers, irrespective of whether an input image was captured by a camera model \emph{known} to our framework.

\subsection{Related Work}
\label{sec:relatedwork}

\subsubsection{Camera Model Attribution.}
Classical approaches to camera model attribution typically construct parametric models of particular physical or algorithmic in-camera processes~\cite{choi2006lens,filler2008using,choi2006jpeg,chen2015camera,bayram2005source,tuama2015source}. Others operate \emph{blindly}, viewing attribution as a texture classification problem, and derive a set of heuristically designed features irrespective of their physical meaning~\cite{marra2017study,kharrazi2004blind,gloe2012feature,xu2012camera}. Recent methods based on deep learning take a data-driven approach, obviating the need for explicit prior domain knowledge~\cite{bayar2018towards,kuzin2018camera,bondi2016first,tuama2016camera,bayar2017design}. A common thread unifying most classical and contemporary approaches is image content suppression (employed as a preprocessing step)~\cite{tuama2015source,marra2015evaluation,filler2008using,gloe2012feature,tuama2016camera,bayar2017design}. That is, it is \emph{a priori} assumed that model-specific image acquisition and processing artifacts are wholly contained in the high-frequency pixel non-uniformity noise (corrupted by in-camera processes) as opposed to the low-frequency image content. Notwithstanding, it has been empirically shown~\cite{IEEE2018} that this preprocessing step is not strictly necessary when the model attribution classifier is a sufficiently deep convolutional neural network (convnet), i.e. training may be performed directly in the spatial domain. Here we focus on deceiving convnet classifiers---with and without image content suppression---which can be considered state-of-the-art.

\subsubsection{Camera Model Anonymization.}
As a direct consequence of the fixation of model attribution methods on model-specific artifacts contained within the pixel non-uniformity noise, model anonymization methods have mostly focused on attenuating or misaligning these high-frequency micro-patterns. Notable approaches include flat-fielding~\cite{gloe2007can,bohme2013counter}, pixel non-uniformity estimation and subtraction~\cite{karakuccuk2015adaptive,dirik2014forensic,bonettini2018fooling}, irreversible forced seam-carving~\cite{dirik2014analysis}, image patch replacement~\cite{entrieri2016patch} and image inpainting~\cite{mandelli2017inpainting}. However, a principal issue with the aforementioned approaches is the detectable absence of model-specific artifacts within the anonymized images~\cite{bonettini2019image}. In contrast, we \emph{transform} the underlying model-specific artifacts of images rather than distorting them.

\subsubsection{Adversarial Examples.}
In image classification, adversarial examples refer to misclassified inputs obtained by applying imperceptible non-random perturbations~\cite{szegedy2013intriguing}. Such attacks are well studied and are typically categorized based on the knowledge available to the adversary (as well as whether the attack causes an untargeted or targeted misclassification). Broadly, white-box attacks require complete knowledge (architecture and parameters) of the classifier to be attacked, whereas black-box attacks require only partial knowledge (obtained by querying the targeted classifier). Similar to other image classification tasks, recent research has shown that model attribution convnets are also extremely vulnerable to adversarial examples, particularly in white-box scenarios~\cite{guera2017counter,marra2018vulnerability,chen2019generative,cozzolino2019spoc}. Nevertheless, in the challenging black-box setting, there are clear issues w.r.t. the transferability of adversarial examples~\cite{marra2018vulnerability,cozzolino2019spoc}. In fact, the apparent lack of transferability has been echoed in other image forensic classification tasks~\cite{barni2019transferability,gragnaniello2018analysis}, such as median filtering and resizing detection, which is in stark contrast to what has been observed in classical object-centric classification tasks.

\subsubsection{Generative Adversarial Networks.}
Generative adversarial networks (GANs) \cite{goodfellow2014generative} offer a viable framework for training generative models and are increasingly being used for tasks such as image generation~\cite{karras2017progressive}, image editing~\cite{zhu2016generative} and representation learning~\cite{mathieu2016disentangling}. Extending this framework to conditional image generation applications, conditional GANs (cGANs)~\cite{mirza2014conditional} have been successfully applied to image-to-image translation~\cite{zhu2017unpaired,isola2017image,zhu2017toward} and modifying image attributes~\cite{perarnau2016invertible,yang2018learning}. Most relevant to our work are the approaches of MISL~\cite{chen2018mislgan,chen2019generative} and SpoC~\cite{cozzolino2019spoc}, which both propose \emph{multiple} camera model anonymizing (unconditional) GANs with access to a single fixed evaluator. That is, for each target camera model, a separate (generator, discriminator) tuple is trained, thus markedly increasing the computational cost. Moreover, MISL only considers the transformation of high-frequency artifacts contained within noise residuals. Differently, SpoC \emph{implicitly} aims to transform image- and noise residual-based artifacts. Using a fixed preprocessor, the discriminative networks (including the evaluator) receive as input the original image concatenated channel-wise with its noise residual. The main problem with concatenating these modalities is that it does not necessarily constrain the discriminative networks to reason over the specific information present in each. Notably, neural networks are known to take \emph{shortcuts}~\cite{geirhos2020shortcut}, and we posit that the discriminative networks learn to rely almost entirely on the noise residual components, as this high-frequency information is not obfuscated by the non-discriminative image content (leading to faster convergence).

\section{Method}
\label{sec:method}

In this section, we outline the attack setting and desiderata, and explain the motivation and core components of our conditional adversarial camera model anonymizer (Cama). Refer to~\Cref{sec:motivation} for an extended motivation.

\subsection{Attack Setting and Desiderata}
We denote by $x\in\mathbb{R}^d$ and $y\in\mathbb{N}_c=\{1,\dots,c\}$ an image and its ground truth (source) camera model label, respectively, sampled from a dataset $p_{\text{data}}$. Consider a \emph{target} (i.e. to be attacked) convnet classifier $F$ with $c$ classes trained over input-output tuples $(x,y)\sim p_{\mathrm{data}}(x,y)$. Given $x$, $F$ outputs a prediction vector of class probabilities $F:x\mapsto F(x)\in[0,1]^{c}$.

In this work, we operate in a \emph{non-interactive black-box setting}: we do not assume to have knowledge of the parameters, architecture or training randomness of $F$, nor can we interact with it. We do, however, assume that we can sample from a dataset similar to $p_{\mathrm{data}}$, which we denote by $q_{\mathrm{data}}$. Precisely, we can sample tuples of the following form: $(x,y)\sim q_{\text{data}}(x,y)$ s.t. $y\in\mathbb{N}_{c^\prime}$, where $c^\prime \leq c$. That is, the set of possible image class labels in $p_{\text{data}}$ is a superset of the set of possible image class labels in $q_{\text{data}}$, i.e. $\mathbb{N}_{c}\supseteq \mathbb{N}_{c^\prime}$.

Suppose $(x,y)\sim q_{\text{data}}(x,y)$ and $y^\prime \in\mathbb{N}_{c^\prime}$, where $y^\prime \neq y$ is a target label. Our aim is to learn a function $G:(x,y^\prime)\mapsto x^\prime \approx x$ s.t. the maximum probability satisfies $\argmax_{i} F(x^\prime)_i=y^\prime$. This is known as a \emph{targeted} attack, whereas the maximum probability of an \emph{untargeted} attack must satisfy $\arg \max_{i} F(x^\prime)_i\neq y$. This work focuses on targeted attacks.

\subsection{Cama: Conditional Adversarial Camera Model Anonymizer}
\label{sec:cama}

\begin{figure}[t]
\centering
\begin{tikzpicture}
[font=\scriptsize\fontfamily{phv}\selectfont]
\node [rectangle, rounded corners=2pt,draw, text width=10pt,minimum width=20pt, text badly centered,minimum height=20pt,  inner sep=0pt,outer sep=0pt, very thick, text=black,fill=Cerulean!80] (inputImage) at (10pt,120pt) {$\text{x}$};

\node [rectangle, rounded corners=2pt, draw, text width=10pt,text badly centered,minimum width=20pt, minimum height=20pt,  inner sep=0pt,outer sep=0pt, very thick, text=black,fill=red!80] (targetLabel) at (10pt,80pt) {$\text{y}^\prime$};

\node [trapezium, rounded corners=2pt,trapezium angle=70, rotate=-90, minimum width=30pt, minimum height=20pt, draw,  very thick, fill=gray!30] (genEnc) at (50pt,120pt) {};

\node [trapezium, rounded corners=2pt,trapezium angle=70, rotate=90, minimum width=30pt, minimum height=20pt, draw,  very thick, fill=gray!30] (genDec) at (70pt,120pt) {};

\node [rectangle,rounded corners=2pt,text badly centered,text width=10pt, minimum width=20pt, minimum height=20pt, inner sep=0pt,outer sep=0pt,  very thick, text=black,draw,fill=TealBlue!80] (fakeImage) at (110pt,120pt) {$\text{x}^\prime$};

\node [trapezium, trapezium angle=70, rotate=-90, minimum width=30pt, minimum height=20pt, draw,  rounded corners=2pt, very thick, fill=gray!30] (enc) at (150pt,120pt) {};

\node [trapezium, rounded corners=2pt,trapezium angle=70, rotate=90, minimum width=30pt, minimum height=20pt, draw,  very thick, fill=gray!30] (dec) at (170pt,120pt) {};

\node [rectangle,rounded corners=2pt,text badly centered,text width=10pt, minimum width=20pt, minimum height=20pt,  draw, very thick, text=black,fill=TealBlue!30]  (highFreq) at (210pt,120pt) {$\text{x}^\prime_\text{H}$};

\node [rectangle,rounded corners=2pt,text badly centered,text width=10pt, minimum width=20pt, minimum height=20pt,  draw, very thick, text=black,fill=TealBlue!30]  (lowFreq) at (210pt,80pt) {$\text{x}^\prime_\text{L}$};

\node [trapezium,rounded corners=2pt, trapezium angle=70, rotate=-90, minimum width=30pt, minimum height=20pt, draw,  very thick, fill=gray!30,text=black] (highClf) at (250pt,120pt) {\rotatebox{90}{$\text{E}_\text{H}$}};

\node [trapezium, rounded corners=2pt,trapezium angle=70, rotate=-90, minimum width=30pt, minimum height=20pt, draw,  very thick, fill=gray!30,text=black] (lowClf) at (250pt,80pt) {\rotatebox{90}{$\text{E}_\text{L}$}};

\node [trapezium,rounded corners=2pt, trapezium angle=70, rotate=-90, minimum width=30pt, minimum height=20pt, draw,  very thick, fill=gray!30,text=black] (disc) at (100pt,80pt) {\rotatebox{90}{$\text{D}$}};

\node [rectangle, rounded corners=2pt, minimum width=40pt,minimum height=15pt, text width=40pt,text badly centered, inner sep=0pt,outer sep=0pt,draw, very thick, fill=yellow] (subtract) at (160pt,80pt) {Subtract};

\node [rectangle, minimum width=30pt, minimum height=10pt,text=black,fill=gray!30] (genText) at (60pt,120pt) {$\text{G}$};

\node [rectangle, minimum width=30pt, minimum height=10pt,text=black,fill=gray!30] (genText) at (160pt,120pt) {$\text{E}_\text{0}$};

\draw [-{Triangle[length=4pt,width=4pt]}] (20pt,85pt) -- (30pt,85pt) -- (30pt,115pt) -- (40pt,115pt) [ very thick];

\draw [-{Triangle[length=4pt,width=4pt]}] (105pt,110pt) -- (105pt,100pt) -- (70pt,100pt) -- (70pt,85pt) -- (90pt,85pt) [ very thick];

\draw [-{Triangle[length=4pt,width=4pt]}] (115pt,110pt) -- (115pt,100pt) -- (150pt,100pt) -- (150pt,88pt) [ very thick];

\draw [-{Triangle[length=4pt,width=4pt]}] (205pt,110pt) -- (205pt,100pt) -- (170pt,100pt) -- (170pt,88pt) [very thick];

 \path
 (inputImage.east) edge[-{Triangle[length=4pt,width=4pt]}, very thick,transform canvas={yshift=5pt}] node [left] {} (genEnc.south)
    (genDec) edge[-{Triangle[length=4pt,width=4pt]}, very thick] node [left] {} (fakeImage)
    (fakeImage) edge[-{Triangle[length=4pt,width=4pt]}, very thick] node [left] {} (enc)
     (dec) edge[-{Triangle[length=4pt,width=4pt]}, very thick] node [left] {} (highFreq)
     (highFreq) edge[-{Triangle[length=4pt,width=4pt]}, very thick] node [left] {} (highClf)
     (subtract.east) edge[-{Triangle[length=4pt,width=4pt]}, very thick] node [left] {} (lowFreq.west)
     (lowFreq.east) edge[-{Triangle[length=4pt,width=4pt]}, very thick] node [left] {} (lowClf.south)
     (targetLabel) edge[-{Triangle[length=4pt,width=4pt]}, very thick,transform canvas={yshift=-5pt}] node [left] {} (disc.south);
\end{tikzpicture}
\caption{Flow diagram of our method Cama given an input tuple $(x,y^\prime)$}
\label{fig:cama}
\end{figure}
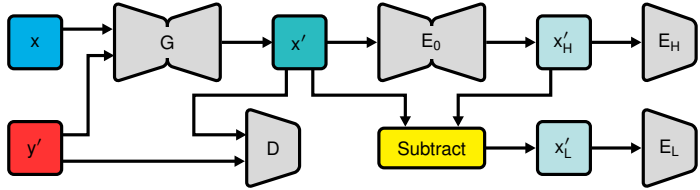

In this framework, our model has two class conditional components: a generator $G$ that transforms an image $x$ conditioned on a target class label $y^\prime$, and a discriminator $D$ that predicts whether the low-level high-frequency pixel value dependency features of any given image conditioned on a label are real or fake. In addition, our model has a fixed (w.r.t. its parameters) dual-stream discriminative decision-making component $E$ (evaluator) that decides whether a transformed image $x^\prime$ belongs to its target class $y^\prime$. In essence, $E$ serves as a \emph{surrogate} for the non-interactive black-box $F$. W.r.t. $E$, a transformed image $x^\prime$ is decomposed into its high and low spatial frequency components ($x^\prime_\mathrm{H}$ and $x^\prime_\mathrm{L}$, respectively), via $E_\mathrm{0}$, with each assigned to a separate stream ($E_\mathrm{H}$ and $E_\mathrm{L}$, respectively). The evaluator then reasons over the information present in $x^\prime_\mathrm{H}$ and $x^\prime_\mathrm{L}$ separately (via $E_\mathrm{H}$ and $E_\mathrm{L}$, respectively). This reinforces the transformation process, as $G$ is constrained to transform both high and low spatial frequency camera model-specific artifacts used by the evaluator for discrimination. Cama is illustrated in~\Cref{fig:cama}.

Our objective contains three types of terms: an \emph{adversarial loss} for matching the distribution of transformed images to the data distribution $q_{\text{data}}$; a \emph{pixel-wise loss} to incentivize the preservation of image content; and a \emph{classification loss} to encourage $G$ to apply transformations that result in transformed images lying in their respective target classes.

\subsubsection{Adversarial Loss.}
To learn plausible conditional transformations, we apply an adversarial loss to $G$. For $G$ and $D$, the training process alternates between $G$ minimizing
\begin{align}
L_{\mathrm{adv}}&= \mathbb{E}_{ \substack{ x \sim q_{\text{data}} (x) \\ y^\prime \sim q_{\text{data}} ({y})} } \left[ \left( D\left( x^\prime, y^\prime \right) - 1\right)^2\right],\label{eq:adv_loss}
\end{align}
and $D$ minimizing
\begin{align}
\begin{split}
L_{\mathrm{dis}}&= \mathbb{E}_{{(x, y) \sim q_{\text{data}} (x,y)}} \left[ \left( D\left({x}, y \right) - 1\right)^2\right]
\\&+
\dfrac{1}{2} \left[ \mathbb{E}_{ \substack{ {x} \sim q_{\text{data}} ({x}) \\ {y^\prime} \sim q_{\text{data}} ({y})} } \left[ D\left(x^\prime, y^\prime \right)^2\right]
+
 \mathbb{E}_{ \substack{ {x} \sim q_{\text{data}} ({x}) \\ {y^\prime} \sim q_{\text{data}} ({y})} } \left[ D\left(x, y^\prime \right)^2\right]
 \right].
\label{eq:disc_loss}
\end{split}
\end{align}
Note that this is the matching-aware~\cite{reed2016generative} least squares formulation~\cite{mao2017least} of the generative adversarial objective \cite{goodfellow2014generative}, which offers increased learning stability, generates higher quality results and encourages $G$ to output images \emph{aligned} with their target labels.

In this work, the first layer of the discriminator is a \emph{constrained} convolutional layer~\cite{bayar2016deep} (originally proposed for image manipulation detection), which learns a set of prediction error filters. Each filter's central value is constrained to equal $-1$, whereas its remaining elements are constrained to sum to unity. In other words, the constrained layer extracts prediction error features (low-level high-frequency pixel value dependency features) by learning a normalized linear combination of the central pixel value based on its local neighborhood~\cite{bayar2017design}. This concurrently serves to suppress image content.

\subsubsection{Pixel-Wise Loss.}
Although $D$ constrains $G$ to learn credible transformations, it does not ensure that an input image's content is preserved. To incentivize this, we incorporate a simple pixel-wise $L^1$ norm loss between $x^\prime$ and ${x}$:
\begin{align}
L_\mathrm{pix}&= \mathbb{E}_{ \substack{ {x} \sim q_{\text{data}} ({x}) \\ {y^\prime} \sim q_{\text{data}} ({y})} }  \left[ \left\lVert {x} - x^\prime \right\rVert_1\right].
\end{align}
The addition of this loss tasks $G$ with producing images that are close to the ground truth input image, i.e. ${x^\prime \approx {x}}$. We prefer an $L^1$ norm loss over an $L^2$ norm loss since it has been observed to encourage less blurring~\cite{zhu2017unpaired}.

\subsubsection{Classification Loss.}
Ideally, $x^\prime$ should possess all relevant target model-specific artifacts that are the result of in-camera processes, i.e. artifacts associated with $y^\prime$. However, the previously introduced adversarial and pixel-wise losses give no guarantees as to whether $x^\prime$ lies in class $y^\prime$ according to an attribution classifier such as $F$. In particular, as $G$ is guided by $D$, the adversarial game may fixate on discovering and fixing peculiarities in $x^\prime$ such as abnormal interpolation patterns~\cite{arandjelovic2019object,chen2019self}. This could result in a failure to transform the \emph{range} of salient model-specific (high and low spatial frequency) artifacts learned by a discriminative classifier, i.e. $\argmax_{i} F(x^\prime)_i \neq y^\prime$. To evaluate and reinforce the transformation process, we adopt a pre-trained fixed dual-stream camera model attribution convnet classifier $E$, where $E$ is employed as a \emph{proxy} for $F$. To decompose an input image $x^\prime$ into its high and low spatial frequency components, the evaluator is prefixed with a preprocessor (convnet) $E_\mathrm{0}:x\mapsto E_\mathrm{0}(x^\prime)=x^\prime_\mathrm{H}$, where $x^\prime_\mathrm{H}$ is the high-frequency noise residual of $x^\prime$. The low spatial frequency components are computed as $x_\mathrm{L}^\prime=x^\prime - x_\mathrm{H}^\prime$. Formally, given $x^\prime$ we propose to reduce the expected negative log-likelihood w.r.t. $y^\prime$ by minimizing
\begin{align}
L_\mathrm{clf}&= -\dfrac{1}{2} \mathbb{E}_{ \substack{ {x} \sim q_{\text{data}} ({x}) \\ {y^\prime} \sim q_{\text{data}} ({y})} } \left[  \log \left( E_\mathrm{H}(x^\prime_\mathrm{H})_{y^\prime}  E_\mathrm{L}(x^\prime_\mathrm{L})_{y^\prime} \right) \right]
,\label{eq:NLL}
\end{align}
where $E_\mathrm{H}$ and $E_\mathrm{L}$ denote the high and low spatial frequency streams of $E$, respectively. Incorporating this loss into the generative objective encourages $G$ to update its parameters s.t. the predicted class of $x^\prime$ is $y^\prime$ according to both streams of $E$.

\subsubsection{Full Objective.}
The full generative objective is as follows:
\begin{align}
L_{\text{gen}}&=L_\mathrm{adv} + \lambda_\mathrm{pix} L_\mathrm{pix} + \lambda_\mathrm{clf} L_\mathrm{clf},\label{eq:fullobj}
\end{align}
where $\lambda_\mathrm{pix}$ and $\lambda_\mathrm{clf}$ are weights that control the contribution of the three objectives. The discriminative objective is as outlined in~\cref{eq:disc_loss}.

\section{Implementation}
To facilitate reproducibility, we make our code publicly available.\footnote{\url{https://github.com/jeroneandrews/cama}}

\setlength{\tabcolsep}{4pt}
\begin{wraptable}{L}{0.48\textwidth}
\centering
\caption{Dataset itemization. Shown are the number of images per class within each set}
\label{table:dataset_split}
\tiny
\begin{tabular}{rlccc}
\toprule
& & \multicolumn{3}{c}{{Set}}\\
\cmidrule(lr){3-5}
 $y$ & {Camera Model} & $q_\mathrm{data}$  & $p_\mathrm{data}$ & $p_\mathrm{test}$ \\
\midrule
 1 & Kodak M1063 & 760 & 765 & 100 \\
 2 & Casio EX-Z150  & 324 & 335  & 100 \\
 3 & Nikon CoolPixS710  & 352   & 330  &  100 \\
 4 & Praktica DCZ5.9  & 345  &  358  & 100 \\
 5 & Olympus mju-1050SW    & 374  & 379 & 100 \\
 6 & Ricoh GX100   & 353 & 296  & 100\\
7 & Rollei RCP-7325XS    & \textendash & 339 & 100 \\
 8 & Panasonic DMC-FZ50    & \textendash & 677 & 100\\
 9 & Samsung NV15    & \textendash & 396 & 100 \\
 10 & Samsung L74wide   & \textendash & 432 & 100\\
 11 & Fujifilm FinePixJ50    & \textendash & 397 & 100\\
 12 & Canon Ixus70    & \textendash & 333 & 100\\
\midrule
 \multicolumn{2}{r}{{Total}} & 2508 & 5037 & 1200\\
\bottomrule
\end{tabular}
\end{wraptable}
\setlength{\tabcolsep}{1.4pt}

\subsubsection{Dataset.}
To provide a comparison on a prevalent digital image forensics benchmark, we use the Dresden image database~\cite{gloe2010dresden} of RGB \texttt{.jpeg} images. Specifically, we use a subset of images from 12 camera models centrally cropped to a common resolution of ${512\times 512}$. We partition the images into three disjoint sets (\Cref{table:dataset_split}), which are disjoint w.r.t. the specific devices used to capture the images. Throughout, sets $q_{\mathrm{data}}$, $p_{\mathrm{data}}$, and $p_{\mathrm{test}}$ are used for constructing non-interactive black-box attacks, target classifier training, and evaluating attack methods, respectively.

\subsubsection{Network Architecture.}
The generative network is adapted from~\cite{johnson2016perceptual} and contains two residual blocks. Following~\cite{chen2018mislgan}, prior to being fed to the network, an input image $x\in\mathbb{R}^d$ is preprocessed (using the Bayer `RGGB' pattern) s.t. it is projected to a color filter array mosaic pattern and then back to a demosaiced RGB image. This serves to remove an input image's original demosaicing traces and forces the generator to \emph{re-demosaic} its input w.r.t. the target camera model condition. The preprocessed image and target label condition $y^\prime\mapsto\{0,1\}^d$ are then concatenated in the channel dimension, where the $y^\prime$-th channel of a target condition is filled with ones with the remaining channels filled with zeros. The discriminative network is adapted from~\cite{isola2017image} and operates at patch-level by classifying $34 \times 34$ overlapping input patches as real or fake. The constrained convolutional layer has three ${5\times5}$ filters with stride and zero-padding equal to 1 and 2, respectively, s.t. the output of this layer retains the spatial size of its input. Similar in principle to~\cite{perarnau2016invertible}, label conditions are reshaped and concatenated in the filter dimension of the output of the first \emph{standard} convolutional layer.  The evaluator's preprocessor $E_\mathrm{0}$ employs the same underlying architecture as $G$, whereas each stream (i.e. $E_\mathrm{H}$ and $E_\mathrm{L}$) uses a ResNet-18 architecture~\cite{he2016deep}.

\subsubsection{Training Details.}
\label{traindetails}
For training, we perform data augmentation by extracting non-overlapping ${64\times64}$ patches from ${512\times512}$ images $\sim q_{\mathrm{data}}$ and using dihedral group $\mathrm{Dih}_4$ transformations. We first train $E_\mathrm{0}$ to approximate ground truth noise residuals obtained through wavelet-based Wiener filtering.\footnote{We refer the reader to~\cite{fridrich2009digital} for details on this denoising filter.} We minimize an $L^2$ norm loss for 90 epochs using Adam~\cite{kingma2014adam} with default parameters, learning rate \num{1e-4}, weight decay \num{5e-4} and batch size 128. Fixing $E_\mathrm{0}$, we separately train $E_\mathrm{H}$ and $E_\mathrm{L}$ to minimize a negative log-likelihood loss. We train both for 90 epochs using SGD with momentum 0.9, initial learning rate 0.1, weight decay \num{5e-4} and batch size 128. Fixing the modules of $E$, we empirically set $\lambda_\mathrm{pix}=10$ and $\lambda_\textrm{clf}=0.01$ in~\cref{eq:fullobj}. We optimize Cama for 200 epochs using the Adam solver with learning rate \num{2e-4}, momentum parameter $\beta_1=0.5$ and batch size 32.

\section{Experiments}
In this section, we provide evidence supporting our main claims: (i) model anonymization requires the transformation of both high and low spatial frequency artifacts, and (ii) better anonymization performance can be obtained by employing an adversarial evaluator that reasons over specific information present in a transformed image's high and low spatial frequency components \emph{separately}.

\subsection{Experimental Setup}
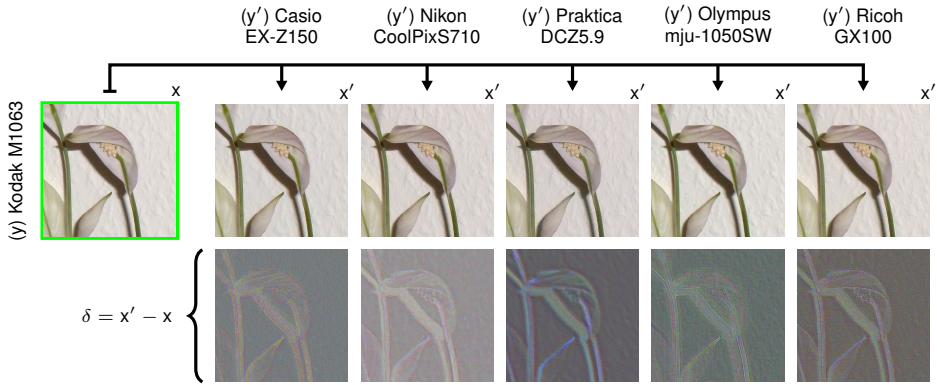
\begin{figure}[t]
\centering
\begin{tikzpicture}
[font=\scriptsize\fontfamily{phv}\selectfont]
\node [rectangle,minimum height=30pt, minimum width=20pt, text badly centered, draw=green,  inner sep=0pt,outer sep=0pt,  line width=2.5pt,  text=black, label={[anchor=center, inner sep=0pt, outer sep=0pt,xshift=-10pt,rotate=90]west:\parbox{55pt}{\centering ($\text{y}$) Kodak M1063}},
label={[anchor=center, inner sep=0pt, outer sep=0pt,yshift=5pt]north east:\parbox{45pt}{\centering $\text{x}$}}] (0) at (15pt,120pt) {\includegraphics[height=50pt]{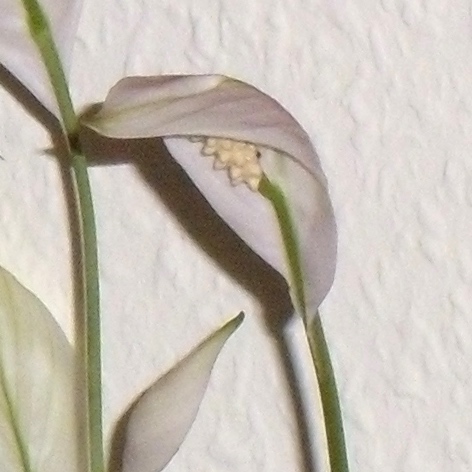}};

\node [rectangle, minimum height=30pt, minimum width=20pt, text badly centered, inner sep=0pt,outer sep=0pt, very thick, text=black, label={[anchor=center, inner sep=0pt, outer sep=0pt,yshift=30pt]north:\parbox{45pt}{\centering ($\text{y}^\prime$) Casio\\EX-Z150}},
label={[anchor=center, inner sep=0pt, outer sep=0pt,yshift=5pt]north east:\parbox{45pt}{\centering $\text{x}^\prime$}}] (1) at (80pt,120pt) {\includegraphics[height=50pt]{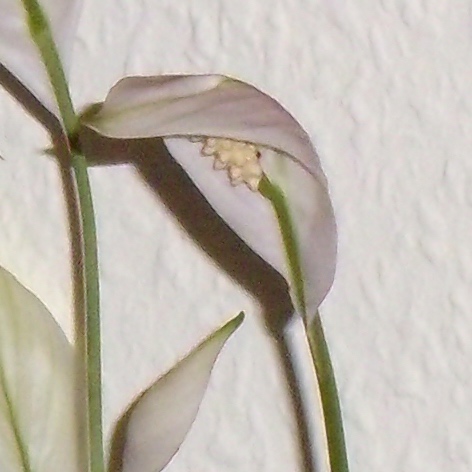}};

\node [rectangle, minimum height=30pt, minimum width=20pt, text badly centered, inner sep=0pt,outer sep=0pt, very thick, text=black, label={[anchor=center, inner sep=0pt, outer sep=0pt,yshift=30pt]north:\parbox{45pt}{\centering ($\text{y}^\prime$) Nikon\\CoolPixS710}},
label={[anchor=center, inner sep=0pt, outer sep=0pt,yshift=5pt]north east:\parbox{45pt}{\centering $\text{x}^\prime$}}] (2) at (135pt,120pt) {\includegraphics[height=50pt]{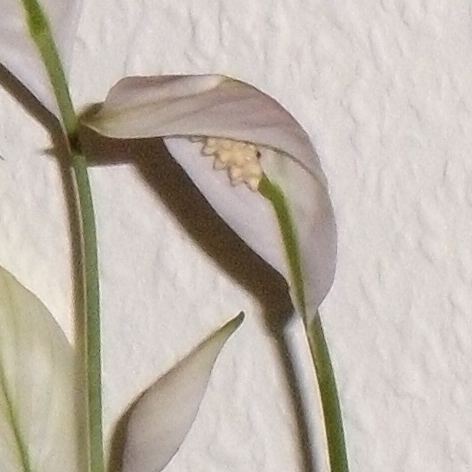}};

\node [rectangle, minimum height=30pt, minimum width=20pt, text badly centered, inner sep=0pt,outer sep=0pt, very thick, text=black, label={[anchor=center, inner sep=0pt, outer sep=0pt,yshift=30pt]north:\parbox{45pt}{\centering ($\text{y}^\prime$) Praktica\\DCZ5.9}},
label={[anchor=center, inner sep=0pt, outer sep=0pt,yshift=5pt]north east:\parbox{45pt}{\centering $\text{x}^\prime$}}] (3) at (190pt,120pt) {\includegraphics[height=50pt]{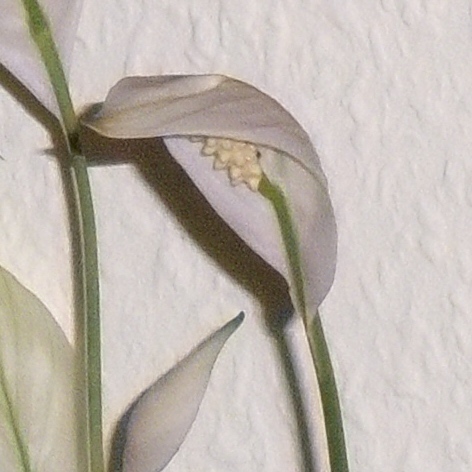}};

\node [rectangle, minimum height=30pt, minimum width=20pt, text badly centered, inner sep=0pt,outer sep=0pt, very thick, text=black, label={[anchor=center, inner sep=0pt, outer sep=0pt,yshift=30pt]north:\parbox{45pt}{\centering ($\text{y}^\prime$) Olympus\\mju-1050SW}},
label={[anchor=center, inner sep=0pt, outer sep=0pt,yshift=5pt]north east:\parbox{45pt}{\centering $\text{x}^\prime$}}] (4) at (245pt,120pt) {\includegraphics[height=50pt]{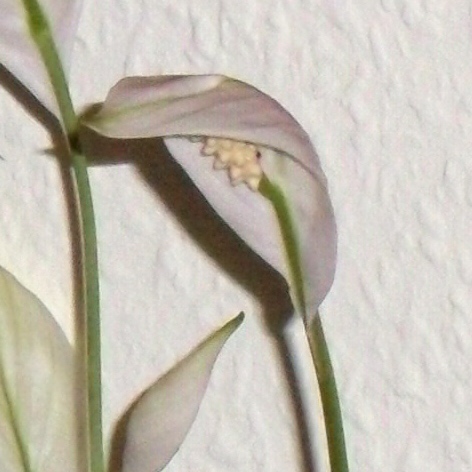}};

\node [rectangle, minimum height=30pt, minimum width=20pt, text badly centered, inner sep=0pt,outer sep=0pt, very thick, text=black, label={[anchor=center, inner sep=0pt, outer sep=0pt,yshift=30pt]north:\parbox{45pt}{\centering ($\text{y}^\prime$) Ricoh\\GX100}},
label={[anchor=center, inner sep=0pt, outer sep=0pt,yshift=5pt]north east:\parbox{45pt}{\centering $\text{x}^\prime$}}] (5) at (300pt,120pt) {\includegraphics[height=50pt]{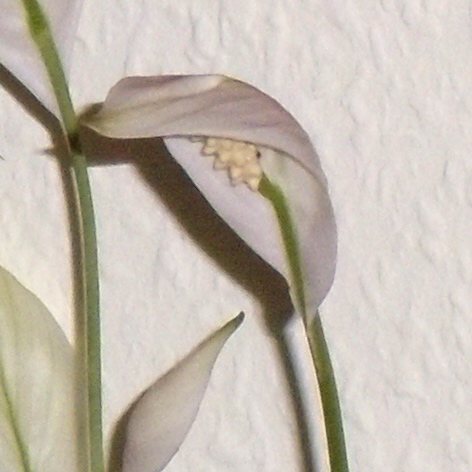}};

\draw [-,very thick] (12.5pt,150pt) -- (17.5pt,150pt)  node[right,yshift=5pt] {};
\draw [-,very thick] (15pt,150pt) -- (15pt,160pt) -- (80pt,160pt) -- (135pt,160pt) -- (190pt,160pt) -- (245pt,160pt) -- (300.6pt,160pt) [];
\draw [-{Triangle[length=4pt,width=4pt]},very thick] (80pt,160pt)  -- (80pt,150pt) node[right,yshift=5pt] {};
\draw [-{Triangle[length=4pt,width=4pt]},very thick] (135pt,160pt) -- (135pt,150pt)node[right,yshift=5pt] {};
\draw [-{Triangle[length=4pt,width=4pt]},very thick] (190pt,160pt) -- (190pt,150pt)node[right,yshift=5pt] {};
\draw [-{Triangle[length=4pt,width=4pt]},very thick] (245pt,160pt) -- (245pt,150pt)node[right,yshift=5pt] {};
\draw [-{Triangle[length=4pt,width=4pt]},very thick] (300pt,160pt) -- (300pt,150pt)node[right,yshift=5pt] {};

\node [rectangle, minimum height=30pt, minimum width=20pt, text badly centered, inner sep=0pt,outer sep=0pt, very thick, text=black] (6) at (80pt,65pt) {\includegraphics[height=50pt]{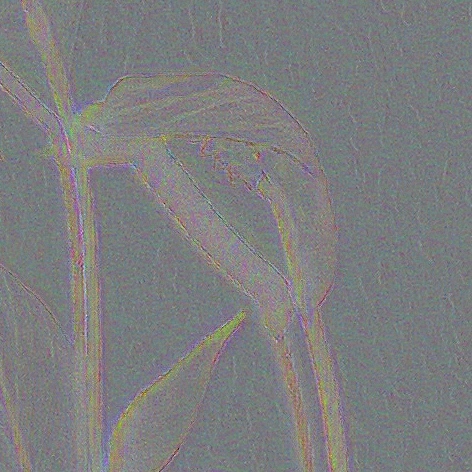}};

\node [rectangle, minimum height=30pt, minimum width=20pt, text badly centered, inner sep=0pt,outer sep=0pt, very thick, text=black] (7) at (135pt,65pt) {\includegraphics[height=50pt]{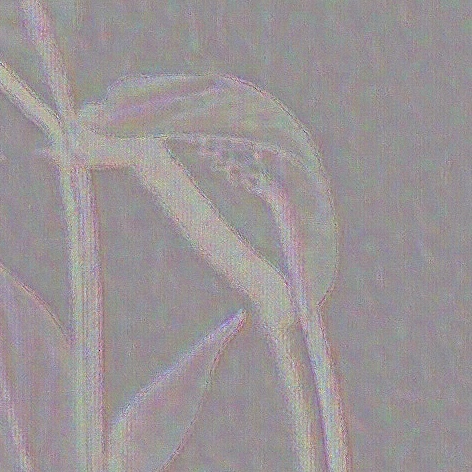}};

\node [rectangle, minimum height=30pt, minimum width=20pt, text badly centered, inner sep=0pt,outer sep=0pt, very thick, text=black] (8) at (190pt,65pt) {\includegraphics[height=50pt]{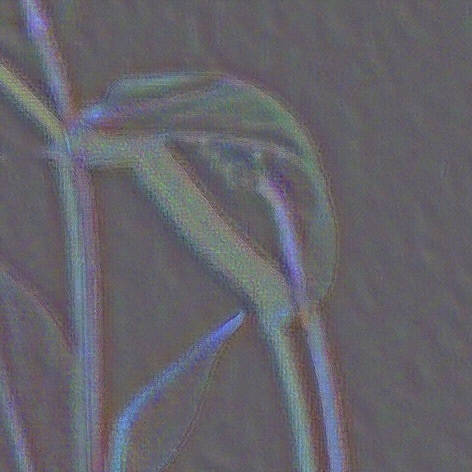}};

\node [rectangle, minimum height=30pt, minimum width=20pt, text badly centered, inner sep=0pt,outer sep=0pt, very thick, text=black] (9) at (245pt,65pt) {\includegraphics[height=50pt]{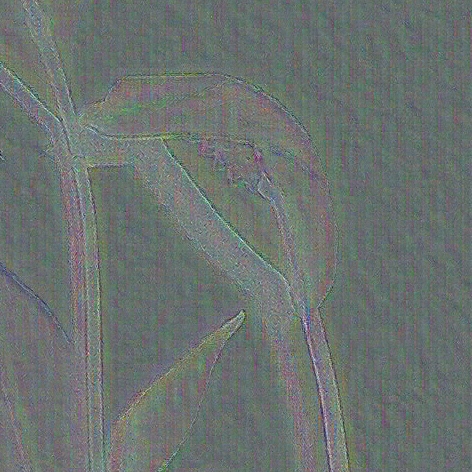}};

\node [rectangle, minimum height=30pt, minimum width=20pt, text badly centered, inner sep=0pt,outer sep=0pt, very thick, text=black] (10) at (300pt,65pt) {\includegraphics[height=50pt]{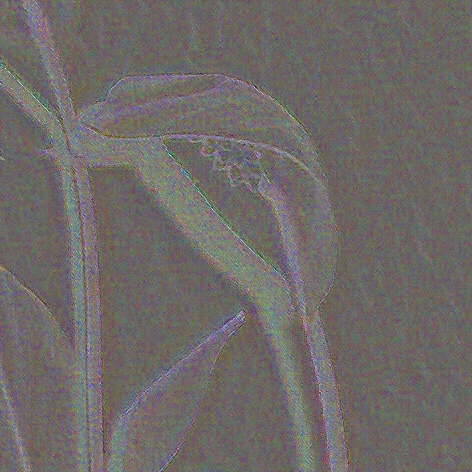}};

\draw [decorate,decoration={brace,amplitude=5pt},yshift=0pt, very thick]
(50pt,40pt) -- (50pt,90pt) node [black,midway, label={[anchor=center, inner sep=0pt, outer sep=0pt,xshift=-25pt]west:\parbox{70pt}{\centering $\mathrm{\delta}=\text{x}^\prime-\text{x}$}}] {};
\end{tikzpicture}
\caption{Example of Cama transformed images $x^\prime$ with different target label conditions $y^\prime$ given an in-distribution input image $x$. The applied transformations (amplified for visualization purposes) are shown as $\delta$}
\label{fig:examples}
\end{figure}

All networks used in this work are fully-convolutional, therefore we can both classify and anonymize images of arbitrary size. Evaluation is always performed on $512\times 512$ images $\sim p_{\textrm{test}}$. \Cref{fig:examples} illustrates Cama's ability to transform an input image using different target label conditions. The applied transformations do not alter the image content and are (largely) imperceptible. See \Cref{extendedqualresults} for additional qualitative results on both in-distribution and out-of-distribution images.

\subsubsection{Baselines.}
To contextualize our approach, we compare against several baselines quantitatively. From the GAN literature~(\Cref{sec:relatedwork}), we recast the unconditional approaches MISL~\cite{chen2018mislgan,chen2019generative} and SpoC~\cite{cozzolino2019spoc} as cGANs. The evaluator of MISL has a single-stream and is prefixed with a constrained convolutional layer. The discriminator and single-stream evaluator of SpoC use a fixed preprocessor, which concatenates an RGB image and its third-order finite differences channel-wise. To ensure a fair comparison, we implement the baselines using the same architecture and details as Cama (where appropriate), and set the parameters of all methods s.t. the mean peak signal-to-noise ratio is approximately 35dB. See~\Cref{extendedresults} for additional comparisons to targeted attack methods from the adversarial examples literature (and mean peak signal-to-noise ratios when anonymizing images $\sim p_{\mathrm{test}}$). We omit these from the main body due to space limitations and their poor performance in contrast to the selected baselines.

\subsubsection{Target Classifiers.}
To validate the efficacy of Cama, we vary the architecture of a target camera model attribution classifier $F$, its prefixed preprocessor and its training data ($\sim p_{\textrm{data}}$). Namely, we consider ResNet-18~\cite{he2016deep}, ResNet-50~\cite{he2016deep}, DenseNet-100~\cite{huang2017densely} and VGG-16~\cite{simonyan2014very} architectures. Regarding the preprocessors, image content suppression is performed by fixed hand-crafted high-pass filtering (HP)~\cite{tuama2016camera}, third-order finite differencing (FD)~\cite{cozzolino2019spoc}, constrained convolution filtering (CC)~\cite{bayar2016deep} or wavelet-based Wiener filtering (WW)~\cite{fridrich2009digital}. We also consider the channel-wise concatenation of an RGB image with its third-order finite differences (RGB+FD)~\cite{cozzolino2019spoc}. No preprocessing is simply denoted by RGB. For training, $F$ is trained over tuples $\sim p_{\mathrm{data}}(x,\mathbb{N}_{6})$ or $p_{\mathrm{data}}(x,\mathbb{N}_{12})$. We refer to the former as a \emph{complete overlap} and the latter as a \emph{partial overlap} of known camera model classes.\footnote{Recall that all anonymization methods are trained on $(x,y)\sim q_{\mathrm{data}}(x,\mathbb{N}_{6})$.} The training details (e.g. training epochs, optimizer, data augmentation, etc.) are the same as used for our evaluator streams (described in \Cref{traindetails}). To confirm that the classifiers $F$ can perform accurate camera model attribution, we compute their classification accuracy on relevant non-anonymized images $\sim p_{\mathrm{test}}$. We obtain a mean accuracy of 99.5\%, which varies by 1.5\% from one target classifier to the next.

\subsubsection{Evaluation Measures.}
We consider the targeted success rate (TSR) for evaluating the anonymization ability of a non-interactive black-box attack, which is defined as the fraction of all possible $x^\prime$ that satisfy $\argmax_{i} F(x^\prime)_i=y^\prime$. For completeness, we also report the untargeted success rate (USR), i.e. the fraction of $x^\prime$ that satisfy $\arg \max_{i} F(x^\prime)_i\neq y$. We report these rates separately for in-distribution images (i.e. captured by camera models \emph{known} to an attack framework) and out-of-distribution images (i.e. captured by camera models \emph{unknown} to an attack framework). The best scores are always shown in boldface.

\subsection{Results}
\subsubsection{Same Architecture Complete Overlap.}
We first analyze the success rates against ResNet-18 target classifiers $F$ trained over $(x,y)\sim p_{\mathrm{data}}(x,\mathbb{N}_6)$. Recall that the evaluators used during adversarial training employ the same underlying architecture. \Cref{table:id-resnet18} shows the results when anonymizing in-distribution images $\sim p_{\mathrm{test}}(x,\mathbb{N}_6)$. Examining the results based on a target classifier's preprocessor, our approach has the highest TSR and USR in 5/6 and 4/6 cases, respectively. \Cref{table:ood-resnet18} shows that our approach has the highest TSR in all cases when anonymizing out-of-distribution images $\sim p_{\mathrm{test}}(x,\mathbb{N}_{12} \setminus \mathbb{N}_6)$.\footnote{Note that $\mathbb{N}_{12} \setminus \mathbb{N}_{6}=\{7,\dots,12\}$.}

\setlength{\tabcolsep}{4pt}
\begin{table}[t]
\centering
\caption{TSR (USR) in the \emph{same architecture complete overlap} setting when anonymizing (a) in-distribution images and (b) out-of-distribution images}
\subfloat[]{
\label{table:id-resnet18}
\tiny
\centering
\begin{tabular}{lrrrrrrr}
\toprule
& \multicolumn{6}{c}{{Preprocessor}} \\ \cmidrule(lr){2-7}
Attack & RGB & RGB+FD & FD & WW & CC & HP & Mean \\
\cmidrule(lr){1-1} \cmidrule(lr){2-7} \cmidrule(lr){8-8}
MISL  & 21.8 (60.0) & 42.6 (79.2) & 65.3 (87.9) & 49.3 (83.1) & 80.1 (85.5) & 52.7 (86.7) & 52.0 (80.4)\\
SpoC  & 64.8 (81.0) & \textbf{94.4} (\textbf{95.4}) & 84.3 (92.6) & 90.8 (94.0) & 75.4 (78.9) & 83.5 (\textbf{94.4}) & 82.2 (89.4) \\
Cama  & \textbf{91.2} (\textbf{96.3}) & 86.8 (87.7) & \textbf{94.2} (\textbf{97.4}) & \textbf{97.3} (\textbf{98.0}) & \textbf{88.1} (\textbf{88.6}) & \textbf{89.7} (92.6) & \textbf{91.2} (\textbf{93.4})\\
\bottomrule
\end{tabular}
}

\subfloat[]{
\label{table:ood-resnet18}
\tiny
\centering
\begin{tabular}{lrrrrrrr}
\toprule
& \multicolumn{6}{c}{{Preprocessor}} \\ \cmidrule(lr){2-7}
Attack & RGB & RGB+FD & FD & WW & CC & HP & Mean \\
\cmidrule(lr){1-1} \cmidrule(lr){2-7} \cmidrule(lr){8-8}
MISL  & 29.2 & 46.4 & 73.0 & 56.1 & 86.9 & 56.2 & 58.0 \\
SpoC  & 80.1 & 97.4 & 90.7 & 96.6 & 91.9 & 92.9 & 91.6\\
Cama  & \textbf{98.5} & \textbf{97.7} & \textbf{94.2} & \textbf{98.1} & \textbf{98.8} & \textbf{97.8} & \textbf{97.5}\\
\bottomrule
\end{tabular}
}
\end{table}
\setlength{\tabcolsep}{1.4pt}

\subsubsection{Same Architecture Partial Overlap.}
In this setting, an attack method is unaware of all possible classes of image learned by target classifiers $F$, which are trained over $(x,y)\sim p_{\mathrm{data}}(x,\mathbb{N}_{12})$. This represents a slightly more realistic scenario. \Cref{table:id-resnet18-exp} shows the results when anonymizing in-distribution images. We achieve the highest TSR and USR in 6/6 and 5/6 cases, respectively. W.r.t. anonymizing out-of-distribution, \Cref{table:ood-resnet18-exp} shows that Cama achieves the highest TSR and USR in all cases.

\setlength{\tabcolsep}{4pt}
\begin{table}[t]
\centering
\caption{TSR (USR) in the \emph{same architecture partial overlap} setting when anonymizing (a) in-distribution images and (b) out-of-distribution images}
\subfloat[]{
\label{table:id-resnet18-exp}
\tiny
\centering
\begin{tabular}{lrrrrrrr}
\toprule
& \multicolumn{6}{c}{{Preprocessor}} \\ \cmidrule(lr){2-7}
Attack & RGB & RGB+FD & FD & WW & CC & HP & Mean \\
\cmidrule(lr){1-1} \cmidrule(lr){2-7} \cmidrule(lr){8-8}
MISL  & 17.7 (70.5) & 12.5 (88.4) & 47.4 (90.5) & 26.2 (93.4) & 20.9 (90.4) & 35.4 (91.6) & 26.7 (87.5)\\
SpoC  & 60.9 (80.6) & 79.6 (\textbf{90.9}) & 83.0 (96.0) & 80.4 (96.0) & 49.0 (86.4) & 47.6 (89.5) & 66.8 (89.9)\\
Cama  & \textbf{81.8} (\textbf{91.3}) & \textbf{80.7} (85.3) & \textbf{94.9} (\textbf{98.3}) & \textbf{94.4} (\textbf{97.1}) & \textbf{82.0} (\textbf{93.5}) & \textbf{82.5} (\textbf{95.3}) & \textbf{86.0} (\textbf{93.5})\\
\bottomrule
\end{tabular}
}

\subfloat[]{
\label{table:ood-resnet18-exp}
\tiny
\centering
\begin{tabular}{lrrrrrrr}
\toprule
& \multicolumn{6}{c}{{Preprocessor}} \\ \cmidrule(lr){2-7}
Attack & RGB & RGB+FD & FD & WW & CC & HP & Mean \\
\cmidrule(lr){1-1} \cmidrule(lr){2-7} \cmidrule(lr){8-8}
MISL  & 16.9 (86.6) & 15.4 (81.3) & 57.8 (91.4) & 30.9 (85.7) & 21.1 (84.0) & 40.9 (92.9) & 30.5 (87.0) \\
SpoC  & 70.6 (92.2) & 89.9 (95.1) & 88.6 (95.6) & 89.1 (96.0) & 61.8 (90.6) & 53.7 (86.9) & 75.6 (92.7)\\
Cama  & \textbf{86.6} (\textbf{95.0}) & \textbf{96.5} (\textbf{98.2}) & \textbf{95.5} (\textbf{98.8}) & \textbf{95.6} (\textbf{99.1}) & \textbf{83.6} (\textbf{95.5}) & \textbf{92.1} (\textbf{97.0}) & \textbf{91.7} (\textbf{97.3}) \\
\bottomrule
\end{tabular}
}
\end{table}
\setlength{\tabcolsep}{1.4pt}

\subsubsection{Architecture Transfer Complete Overlap.}
To investigate whether the attacks transfer to other architectures---i.e. architectures that are distinct from an attack method's ResNet-18-based evaluator---we ran experiments using different target classifier architectures: ResNet-50 (R-50), DenseNet-100 (D-100) and VGG-16 (V-16). \Cref{table:id-others} shows the results when anonymizing in-distribution images: the success rates and trend are similar to what we observed when attacking ResNet-18 target classifiers (\Cref{table:ood-resnet18}). In particular, Cama has the highest TSR and USR in 18/18 and 14/18 cases, respectively. On out-of-distribution images, we attain the highest success rate in all cases (\Cref{table:ood-others}).

\setlength{\tabcolsep}{4pt}
\begin{table}[t]
\centering
\caption{TSR (USR) in the \emph{architecture transfer complete overlap} setting when anonymizing (a) in-distribution images and (b) out-of-distribution images}
\subfloat[]{
\label{table:id-others}
\tiny
\centering
\begin{tabular}{lrrrrrrrr}
\toprule
& &\multicolumn{6}{c}{{Preprocessor}} \\ \cmidrule(lr){3-8}
 & Attack & RGB & RGB+FD & FD & WW & CC & HP & Mean \\
\cmidrule(lr){2-2} \cmidrule(lr){3-8} \cmidrule(lr){9-9}
\multirow{3}{*}{\rotatebox[origin=c]{90}{{R-50}}} & MISL  & 20.8 (57.2) & 20.8 (76.0) & 71.3 (89.6) & 55.1 (86.4) & 45.5 (67.8) & 69.7 (\textbf{90.6}) & 47.2 (77.9) \\
& SpoC  & 60.0 (75.6) & 78.7 (\textbf{91.6}) & 90.5 (\textbf{94.5}) & 88.1 (91.6) & 73.1 (81.0) & 77.2 (89.3) & 77.9 (87.3) \\
& Cama  & \textbf{87.9} (\textbf{92.3}) & \textbf{79.8} (83.7) & \textbf{92.0} (92.5) & \textbf{95.1} (\textbf{95.9}) & \textbf{93.9} (\textbf{95.9}) & \textbf{89.3} (89.9) & \textbf{89.7} (\textbf{91.7}) \\

\midrule
\multirow{3}{*}{\rotatebox[origin=c]{90}{{D-100}}} & MISL  & 28.7 (59.1) & 57.4 (75.4) & 62.0 (86.8) & 54.9 (83.6) & 96.8 (\textbf{99.0}) & 62.9 (88.7) & 60.4 (82.1) \\
& SpoC  & 62.0 (77.0) & 86.9 (91.5) & 88.7 (94.2) & 86.1 (92.9) & 78.9 (88.5) & 87.0 (95.9) & 81.6 (90.0) \\
& Cama  & \textbf{97.8} (\textbf{98.7}) & \textbf{97.4} (\textbf{97.8}) & \textbf{98.8} (\textbf{99.3}) & \textbf{99.5} (\textbf{99.8}) & \textbf{98.6} (98.9) & \textbf{96.9} (\textbf{99.5}) & \textbf{98.2} (\textbf{99.0}) \\

\midrule
\multirow{3}{*}{\rotatebox[origin=c]{90}{{V-16}}} & MISL  & 16.6 (83.2) & 60.8 (84.3) & 50.3 (83.9) & 42.0 (83.7) & 56.8 (77.8) & 66.2 (86.2) & 48.8 (83.2) \\
& SpoC  &  46.0 (80.1) & 95.5 (96.8) & 90.3 (95.3) & 69.6 (88.0) & 89.2 (93.3) & 84.5 (91.3) & 79.2 (90.8)\\
& Cama  & \textbf{88.0} (\textbf{95.9}) & \textbf{99.0} (\textbf{99.3}) & \textbf{98.1} (\textbf{98.8}) & \textbf{87.4} (\textbf{96.6}) & \textbf{98.7} (\textbf{99.3}) & \textbf{98.4} (\textbf{98.8}) & \textbf{94.9} (\textbf{98.1}) \\
\bottomrule
\end{tabular}
}

\subfloat[]{
\label{table:ood-others}
\tiny
\centering
\centering
\begin{tabular}{lrrrrrrrr}
\toprule
& &\multicolumn{6}{c}{{Preprocessor}} \\ \cmidrule(lr){3-8}
 & Attack & RGB & RGB+FD & FD & WW & CC & HP & Mean \\
\cmidrule(lr){2-2} \cmidrule(lr){3-8} \cmidrule(lr){9-9}
\multirow{3}{*}{\rotatebox[origin=c]{90}{{R-50}}} & MISL  & 30.3 & 27.8 & 77.1 & 64.9 & 51.4 & 74.8  & 54.4\\
& SpoC  & 75.2 & 83.8 & 95.5 & 96.6 & 87.7 & 86.7 & 87.6\\
& Cama  & \textbf{95.9} & \textbf{96.7} & \textbf{99.2} & \textbf{99.4} & \textbf{98.9} & \textbf{99.1} & \textbf{98.2}\\
\midrule
\multirow{3}{*}{\rotatebox[origin=c]{90}{{D-100}}} & MISL  & 34.9 & 73.3 & 67.6 & 61.4 & 97.5 & 66.9 & 66.9 \\
& SpoC  & 80.9 & 93.6 & 94.2 & 90.9 & 84.4 & 92.5 & 89.4\\
& Cama  & \textbf{99.1} & \textbf{99.6} & \textbf{99.6} & \textbf{99.7} & \textbf{99.3} & \textbf{98.6} & \textbf{99.3} \\
\midrule
\multirow{3}{*}{\rotatebox[origin=c]{90}{{V-16}}} & MISL  & 17.1 & 71.9 & 50.6 & 50.8 & 68.3 & 71.6 & 55.0 \\
& SpoC  & 55.8 & 98.2 & 92.3 & 77.4 & 96.3 & 93.8  & 85.6 \\
& Cama  & \textbf{91.6} & \textbf{99.8} & \textbf{98.2} & \textbf{91.7} & \textbf{99.3} & \textbf{99.3} & \textbf{96.6}  \\
\bottomrule
\end{tabular}
}
\end{table}
\setlength{\tabcolsep}{1.4pt}

\subsubsection{Architecture Transfer Partial Overlap.}
This represents the most realistic and interesting setting for assessing the performance of a non-interactive black-box attack. Notably, on in-distribution images we achieve the highest TSR and USR in 18/18 and 15/18 cases (\Cref{table:id-others-exp}), respectively. Moreover, as shown in \Cref{table:ood-others-exp}, we attain the best success rates in every case when anonymizing out-of-distribution images.

\setlength{\tabcolsep}{4pt}
\begin{table}[t]
\centering
\caption{TSR (USR) in the \emph{architecture transfer partial overlap} setting when anonymizing (a) in-distribution images and (b) out-of-distribution images}
\subfloat[]{
\label{table:id-others-exp}
\tiny
\begin{tabular}{lrrrrrrrr}
\toprule
& &\multicolumn{6}{c}{{Preprocessor}} \\ \cmidrule(lr){3-8}
 & Attack & RGB & RGB+FD & FD & WW & CC & HP & Mean \\
\cmidrule(lr){2-2} \cmidrule(lr){3-8} \cmidrule(lr){9-9}
\multirow{3}{*}{\rotatebox[origin=c]{90}{{R-50}}} & MISL  & 8.4 (81.5) & 8.2 (89.5) & 20.8 (91.7) & 19.1 (96.2) & 33.8 (83.0) & 36.5 (\textbf{91.6}) & 21.1 (88.9) \\
& SpoC  & 62.7 (90.3) & 66.4 (\textbf{95.7}) & 73.9 (95.0) & 71.2 (96.6) & 70.9 (94.7) & 51.2 (\textbf{91.6}) & 66.0 (94.0) \\
& Cama  & \textbf{92.6} (\textbf{97.6}) & \textbf{73.3} (88.2) & \textbf{83.3} (\textbf{98.2}) & \textbf{91.9} (\textbf{99.3}) & \textbf{81.9} (\textbf{96.4}) & \textbf{75.3} (88.4) & \textbf{83.0} (\textbf{94.7}) \\
\midrule
\multirow{3}{*}{\rotatebox[origin=c]{90}{{D-100}}} & MISL  & 11.5 (71.9) & 31.0 (91.8) & 37.6 (96.9) & 20.6 (94.7) & 16.2 (93.1) & 24.1 (98.3) & 23.5 (91.1) \\
& SpoC  & 52.6 (81.9) & 89.2 (96.2) & 85.9 (97.5) & 79.8 (94.5) & 76.3 (96.5) & 60.9 (94.5) & 74.1 (93.5)  \\
& Cama  & \textbf{88.3} (\textbf{95.0}) & \textbf{96.7} (\textbf{98.6}) & \textbf{96.3} (\textbf{99.5}) & \textbf{96.9} (\textbf{99.2}) & \textbf{88.9} (\textbf{97.4}) & \textbf{88.9} (\textbf{99.2}) & \textbf{92.7} (\textbf{98.2}) \\
\midrule
\multirow{3}{*}{\rotatebox[origin=c]{90}{{V-16}}} & MISL  & 30.1 (86.5) & 21.5 (96.2) & 46.6 (85.8) & 27.5 (97.2) & 96.9 (\textbf{99.9}) & 28.6 (96.4) & 41.9 (93.7) \\
& SpoC  & 81.7 (93.3) & 81.7 (96.3) & 76.3 (92.4) & 80.5 (95.1) & 64.1 (90.6) & 61.3 (96.2) & 74.3 (94.0)  \\
& Cama  & \textbf{98.3} (\textbf{99.5}) & \textbf{95.1} (\textbf{98.4}) & \textbf{94.5} (\textbf{99.0}) & \textbf{94.3} (\textbf{99.4}) & \textbf{97.3} (98.9) & \textbf{92.9} (\textbf{99.2}) & \textbf{95.4} (\textbf{99.1})\\
\bottomrule
\end{tabular}
}

\centering
\subfloat[]{
\label{table:ood-others-exp}
\tiny
\begin{tabular}{lrrrrrrrr}
\toprule
& &\multicolumn{6}{c}{{Preprocessor}} \\ \cmidrule(lr){3-8}
 & Attack & RGB & RGB+FD & FD & WW & CC & HP & Mean \\
\cmidrule(lr){2-2} \cmidrule(lr){3-8} \cmidrule(lr){9-9}
\multirow{3}{*}{\rotatebox[origin=c]{90}{{R-50}}} & MISL  & 7.6 (79.3) & 9.8 (85.9) & 27.1 (82.2) & 22.8 (85.4) & 39.2 (86.0) & 42.0 (91.9) & 24.8 (85.1) \\
& SpoC  & 70.7 (93.0) & 77.0 (92.8) & 82.8 (94.6) & 80.3 (94.3) & 72.5 (91.3) & 56.7 (94.6) & 73.3 (93.4) \\
& Cama  & \textbf{96.1} (\textbf{97.9}) & \textbf{88.4} (\textbf{94.4}) & \textbf{88.7} (\textbf{96.0}) & \textbf{97.5} (\textbf{99.1}) & \textbf{81.3} (\textbf{93.6}) & \textbf{86.8} (\textbf{97.1}) & \textbf{89.8} (\textbf{96.4})\\
\midrule
\multirow{3}{*}{\rotatebox[origin=c]{90}{{D-100}}} & MISL  & 10.6 (78.4) & 34.7 (85.5) & 40.5 (82.9) & 20.7 (85.6) & 15.1 (80.0) & 29.4 (86.9) & 25.2 (83.2) \\
& SpoC  &  66.5 (89.7) & 93.4 (97.0) & 91.1 (96.5) & 86.9 (96.2) & 82.1 (91.3) & 75.5 (94.2) & 82.6 (94.1) \\
& Cama  & \textbf{92.3} (\textbf{95.5}) & \textbf{97.8} (\textbf{98.5}) & \textbf{95.2} (\textbf{98.3}) & \textbf{97.3} (\textbf{98.8}) & \textbf{91.2} (\textbf{95.6}) & \textbf{94.7} (\textbf{98.0}) & \textbf{94.8} (\textbf{97.4})\\
\midrule
\multirow{3}{*}{\rotatebox[origin=c]{90}{{V-16}}} & MISL  & 33.9 (87.7) & 31.4 (84.6) & 49.8 (88.0) & 23.9 (83.6) & 98.0 (99.6) & 30.6 (83.2) & 44.6 (87.8)  \\
& SpoC  & 77.7 (95.7) & 85.8 (94.9) & 79.0 (92.0) & 84.9 (94.8) & 74.1 (90.6) & 70.0 (87.0) & 78.6 (92.5) \\
& Cama  & \textbf{98.9} (\textbf{99.6}) & \textbf{97.6} (\textbf{98.8}) & \textbf{93.0} (\textbf{97.4}) & \textbf{95.4} (\textbf{97.5}) & \textbf{96.4} (\textbf{98.8}) & \textbf{95.9} (\textbf{97.9}) & \textbf{96.2} (\textbf{98.3})\\
\bottomrule
\end{tabular}
}
\end{table}
\setlength{\tabcolsep}{1.4pt}

\subsection{Discussion}
The experimental results validate that our approach (Cama) is able to reliably perform targeted transformations. Importantly, not only can we successfully  perform targeted transformations on in-distribution images captured by camera models \emph{known} to our framework, but also on out-of-distribution images captured by camera models \emph{unknown} to our framework. Most significantly, our attack methodology transfers across different target classifiers, i.e. as we vary a target classifier's architecture, preprocessing module and training data. Our results are non-trivial: for instance, there is no reason that the feature space of a VGG-16-based target classifier should \emph{behave} in the same manner as an adversarial evaluator's ResNet-18. This shows that our method has good generalization ability and that the applied transformations go above and beyond mere adversarial noise. This last point is especially apparent when one considers the results attained using adversarial example methods (\Cref{extendedresults}).

As hypothesized, it is critical to transform both high and low spatial frequency artifacts. This can be readily seen by the poor performance of MISL, which wholly focuses on the transformation of high-frequency artifacts, and therefore cannot attend to lower-frequency model-specific artifacts used by RGB-based target classifiers (i.e. without image content suppression). Moreover, while the method of prediction-error filtering, using a constrained convolutional layer, has been successfully used for camera model attribution, it was originally proposed for image manipulation detection~\cite{bayar2016deep}. It is unclear how these features relate to model-specific artifacts extracted by other methods that suppress image content as a preprocessing step, i.e. other than being useful for image manipulation detection. Principally, when employed by an evaluator (as is the case for MISL), the learned generator is incapable of reliably causing targeted misclassifications when faced by target classifiers that employ a dissimilar image content suppressor and/or architecture.

During optimization, the generator of SpoC is guided by an evaluator and discriminator that concatenate an RGB input image channel-wise with its image residual. We posit that concatenating these input modalities does not effectively force the generative model to update its parameters s.t. lower-frequency artifacts contained in the RGB channels of an input are attended to. As evidenced by the results, the evaluator and discriminator of SpoC pay more attention to the image residual channels. In contradistinction to SpoC, Cama \emph{constrains} its generator using a dual-stream evaluator that independently reasons over high and low spatial frequency artifacts. This consistently improves performance, since the generator is tasked with fooling both streams such that it cannot easily take a \emph{shortcut}---i.e. predominantly focus on the salient high-frequency information. Patently, anonymization methods that are capable of transforming model-specific artifacts of a high and low spatial frequency are better able to deceive unknown non-interactive black-box target classifiers of varying types. This is particularly useful, since we do not know \emph{a priori} in which space a target classifier operates.

\section{Conclusion}
The method proposed in this paper, Cama, offers a way to preserve privacy by transforming an image's ground truth camera model-specific artifacts to those of a disparate target camera model. By formulating the learning procedure as necessitating the transformation of both high and low spatial frequency artifacts, we proposed to incorporate a fixed pre-trained dual-stream evaluator into the generative objective. The evaluator serves to reinforce the transformation process by \emph{independently} reasoning over information present in the high and low spatial frequencies. Experimental results demonstrate that our approach (i) can successfully anonymize images captured by camera models \emph{known} and \emph{unknown} to our framework, and (ii) results in targeted transformations that are non-interactive black-box target classifier-agnostic (i.e. as we vary the architecture, training data and preprocessing module of a target classifier).

While the preservation of privacy is evidently beneficial to certain vulnerable individuals, anonymization could equally be open to misuse~\cite{chesney2018deep}. As society is at present afflicted by the deliberate dissemination of misinformation, we require more robust and reliable digital forensic methods for authenticating the origin and integrity of images. In particular, when faced by synthesized photo-realistic \emph{deepfakes}~\cite{caldwell2020ai}, which additionally mimic the intrinsic artifacts of a target camera model (and/or device) as made possible by Cama.

\subsubsection{Acknowledgments.}
JTAA is supported by the Royal Academy of Engineering (RAEng) and the Office of the Chief Science Adviser for National Security under the UK Intelligence Community Postdoctoral Fellowship Programme.

\clearpage
%
%
\bibliographystyle{splncs04}
\bibliography{egbib}

\clearpage
\appendix
\section{Extended Motivation}
\label{sec:motivation}
The central premise of this work is that there exist both \emph{high} and \emph{low} spatial frequency artifacts that constitute useful signals for standard camera model attribution (and ipso facto camera model anonymization). We now briefly provide discussion in support of this.

We denote by $x\in\mathbb{R}^d$ an observable digital image (e.g. stored as a \texttt{.jpeg} or \texttt{.tiff} file). The in-camera processing pipeline in digital cameras is complex in nature and can vary greatly between camera models. To capture the \emph{common} elements of in-camera processing,~\cite{chen2008determining} proposed the following simplified imaging sensor output model (prior to color interpolation):
\begin{align}
x&=g^\gamma\left[ (1+\Upsilon)u + \Lambda \right]^\gamma + \Theta_q \\
&= (gu)^\gamma\left[1+(\Upsilon+\Lambda/u)\right]^\gamma + \Theta_q,\label{eq:simp_model}
\end{align}
where $u$ is the incident light intensity, $g$ is the color channel gain, $\gamma$ is the gamma correction, $\Upsilon$ is the device-specific zero-mean noise-like factor responsible for photo-response non-uniformity, $\Theta_q$ is the quantization noise and $\Lambda$ is a combination of all other noise sources. Retaining only the first two terms of the Taylor expansion of $\left[1+(\Upsilon+\Lambda/u)\right]^\gamma$ at $\Upsilon+\Lambda/u=0$, \cref{eq:simp_model} becomes:
\begin{align}
x&\approx (gu)^\gamma (1+\gamma\Upsilon + \gamma\Lambda/u) + \Theta_q \\
&={x_0} + {x_0}\Omega + \Theta,
\end{align}
where $x_0=(gu)^\gamma$ is the image content (ideal noise-free sensor response), $\Omega=\gamma\Upsilon$ is the gamma corrected photo-response non-uniformity factor and $\Theta=\gamma x_0 \Lambda/u + \Theta_q$ is a combination of random independent noise sources.

To improve the signal-to-noise between the photo-response non-uniformity signal $x_0\Omega$ and the observable image $x$, the image content $x_0$ is suppressed via a denoising filter $H:x\mapsto \hat{x}_0$:
\begin{align}
x_\mathrm{H}&=x-H(x)\\
&= {x_0} + {x_0}\Omega + \Theta - \hat{x}_0\\
&= (x_0-\hat{x}_0)+(x_0-x)\Omega+x\Omega+\Theta\\
&= x\Omega+\Xi,\label{eq:noise_res}
\end{align}
where $\Xi=\Theta+(x_0-\hat{x}_0)+(x_0-x)\Omega$. Working with the high-frequency noise residual $x_\mathrm{H}$, it is easier to estimate (and detect) $\Omega$, because the image content has been significantly suppressed. Appositely, the unique device-specific factor $\Omega$ is commonly employed for model attribution~\cite{tuama2015source,marra2015evaluation,filler2008using,gloe2012feature,tuama2016camera,bayar2017design}, since it also contains non-unique artifacts (unless corrected for).

As aforementioned (\Cref{sec:intro}), subsequent to the imaging sensor, the effects of $\Omega$ propagate nonlinearly through the in-camera processing steps that result in an observable image $x$ and thus end up also depending on model-specific aspects, such as color interpolation, on-sensor signal transfer, sensor design and compression. Therefore, $x_\mathrm{H}$ contains model-specific artifacts with high spatial frequency, i.e. \emph{features} that change considerably in intensity over short spatial distances. However, due to denoising, $x\Omega$ in \cref{eq:noise_res} no longer corresponds to the photo-response non-uniformity signal, but instead the pixel non-uniformity signal (\cref{fig:pattern_noise}). The attenuation of the image content, through denoising, results in the removal of the low spatial frequency components inherent to photo-response non-uniformity that, for instance, stem from the model-specific camera optics. Therefore, anonymizing $x$ based on the information contained in $x_\mathrm{H}$ precludes the anonymization of discriminative low spatial frequency artifacts contained in $x_\mathrm{L}=x-x_\mathrm{H}$. Remarkably, this crucial point has been overlooked by previous model anonymization works.

\section{Extended Qualitative Results}
\label{extendedqualresults}
In this section, we provide additional figures showing the ability of Cama to transform in-distribution (\Cref{fig:moreexamples}) and out-of-distribution (\Cref{fig:moreexamples2}) input images using different target label conditions.

\begin{figure}[t]
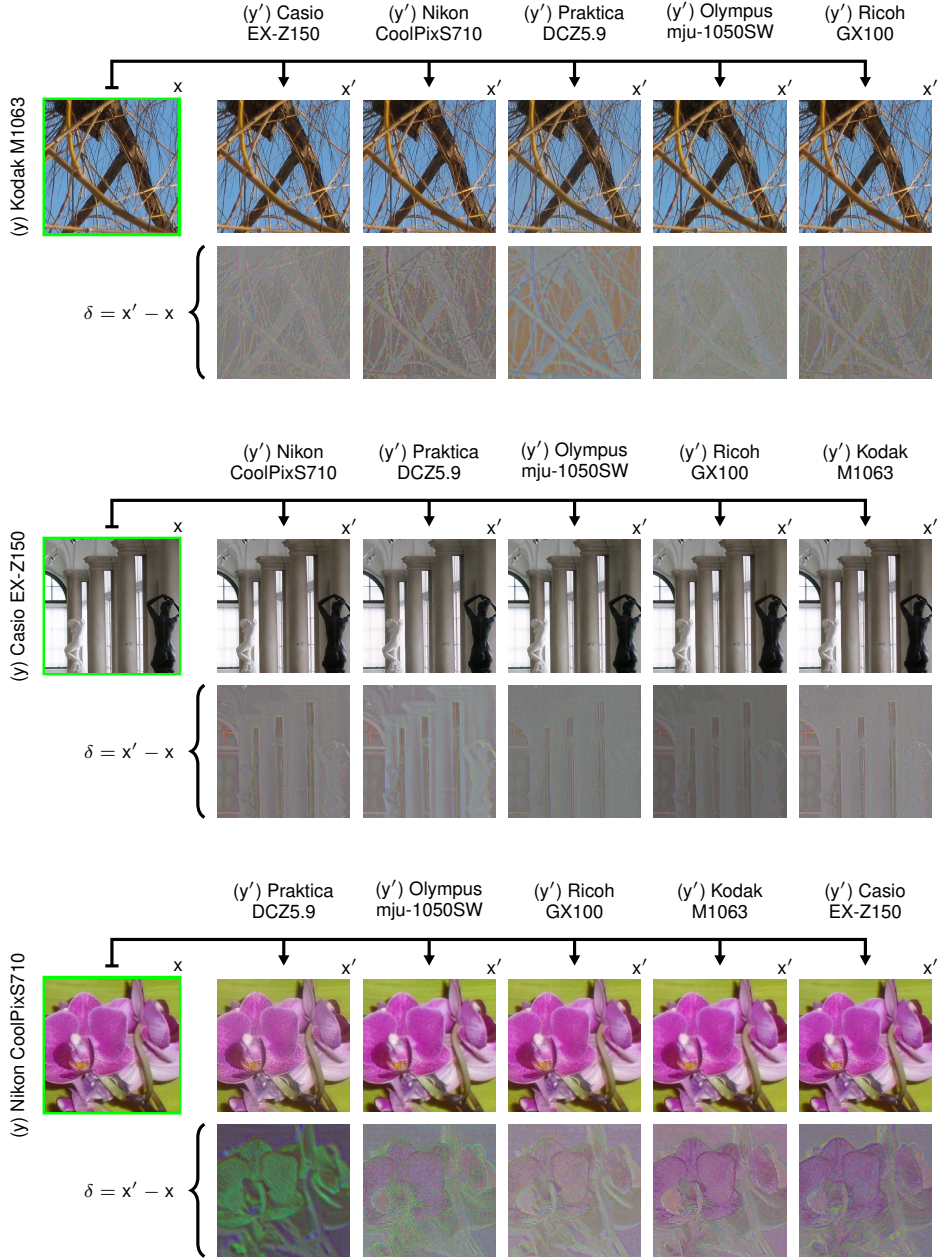

\centering


\caption{Example of Cama transformed images $x^\prime$ with different target label conditions $y^\prime$ given an in-distribution input image $x$. The applied transformations (amplified for visualization purposes) are shown as $\delta$. (Figure continues)}
\label{fig:moreexamples}
\end{figure}

\begin{figure}[t]
\ContinuedFloat
\centering

%
\caption{(Continued). Example of Cama transformed images $x^\prime$ with different target label conditions $y^\prime$ given an in-distribution input image $x$. The applied transformations (amplified for visualization purposes) are shown as $\delta$}
\end{figure}

\begin{landscape}
\begin{figure}[t]
\centering
%

\caption{Example of Cama transformed images $x^\prime$ with different target label conditions $y^\prime$ given an out-of-distribution input image $x$. The applied transformations (amplified for visualization purposes) are shown as $\delta$. (Figure continues)}
\label{fig:moreexamples2}
\end{figure}

\end{landscape}

\begin{landscape}
\begin{figure}[t]
\ContinuedFloat
\centering
%

\caption{(Continued). Example of Cama transformed images $x^\prime$ with different target label conditions $y^\prime$ given an out-of-distribution input image $x$. The applied transformations (amplified for visualization purposes) are shown as $\delta$. (Figure continues)}
\end{figure}

\end{landscape}

\begin{landscape}
\begin{figure}[t]
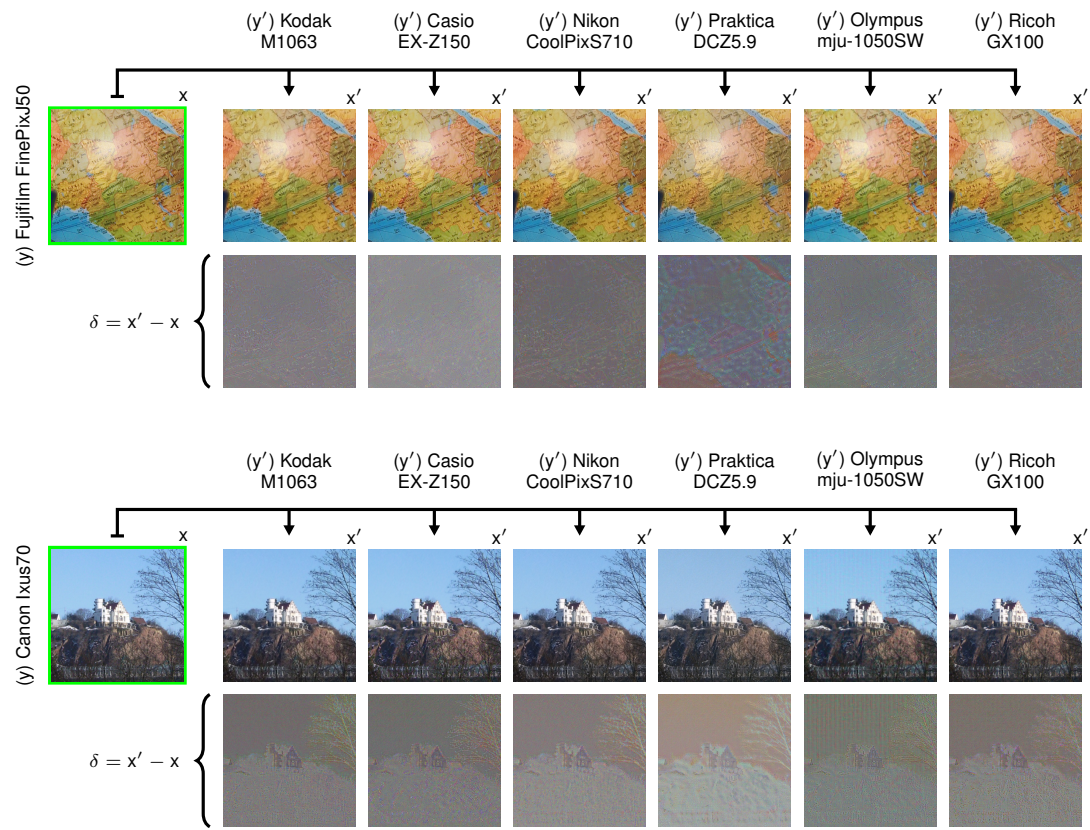

\ContinuedFloat
\centering
%

\caption{(Continued). Example of Cama transformed images $x^\prime$ with different target label conditions $y^\prime$ given an out-of-distribution input image $x$. The applied transformations (amplified for visualization purposes) are shown as $\delta$}
\end{figure}

\end{landscape}

\section{Extended Quantitative Results}
\label{extendedresults}
From the adversarial examples literature, we perform targeted attacks using the following representative approaches: fast gradient sign method (FGSM) \cite{goodfellow2014explaining}, projected gradient descent (PGD) \cite{madry2017towards} and decoupled direction and norm (DDN) \cite{rony2019decoupling}, where the perturbations are constructed using an evaluator $E$ with a ResNet-18 backbone architecture. (See individual descriptions below). Note that we tested several different preprocessing modules, i.e. with and without image content suppression, however results remained roughly commensurate. We therefore utilized the dual-stream evaluator of our approach (as outlined in \Cref{sec:method}).

To ensure a fair comparison, we set the parameters for each method s.t. the peak signal-to-noise ratio is approximately 35dB, except for DDN. DDN is designed to efficiently find small perturbations that fool a proxy model. \Cref{table:psnrs} shows the mean peak signal-to-noise ratio when anonymizing in-distribution images and out-of-distribution images $\sim p_{\mathrm{test}}$, whereas \Cref{table:id-resnet18-add,table:ood-resnet18-add,table:id-resnet18-exp-add,table:ood-resnet18-exp-add,table:id-others-add,table:ood-others-add,table:id-others-exp-add,table:ood-others-exp-add} (below) show the targeted and untargeted success rates of these attack methods in comparison to our approach Cama.

\subsubsection{FGSM.}
For an $L^\infty$-bounded adversary, the one-step FGSM computes a targeted adversarial example as:
\begin{align}
x^\prime=x-\epsilon\sign(-\nabla_x \log E(x)_{y^\prime}),
\end{align}
 where $\epsilon$ is the magnitude of the perturbation.

\subsubsection{PGD.}
PGD is an iterative variant of FGSM. At iteration $t+1$, PGD follows the update rule:
\begin{align}
x^\prime_{t+1}=\proj(x^\prime_{t}-\alpha\sign(-\nabla_x \log E(x^\prime_{t})_{y^\prime})),
\end{align}
where $\alpha$ is the step size and $\proj$ is a projection into the $L^\infty$-ball with radius $\epsilon$ and center $x$.

\subsubsection{DDN.}
For an $L^2$-bounded adversary, DDN is an iterative method (similar to PGD) that induces a targeted misclassification with low $L^2$ norm. At iteration $t+1$, DDN follows the update rule:
\begin{align}
x^\prime_{t+1}=\proj(x^\prime_{t}-\alpha \frac{-\nabla_x\log E(x^\prime_{t})_{y^\prime}}{\lVert -\nabla_x\log E(x^\prime_{t})_{y^\prime} \rVert_2}),
\end{align}
where $\alpha$ is the step size and $\proj$ is a projection into the $\ell_2$-ball with radius $\epsilon_{t+1}$  and center $x$. The radius is adapted based on the current distortion s.t. if $x^\prime_{t}$ is an adversarial example then $\epsilon_{t+1}=(1-\gamma)\lVert x^\prime_{t} -x\rVert_2$, otherwise $\epsilon_{t+1}=(1+\gamma)\lVert x^\prime_{t} -x\rVert_2$, where $\gamma$ is a factor to modify the norm in each iteration.

\setlength{\tabcolsep}{4pt}
\begin{table}[t]
\centering
\caption{Mean peak signal-to-noise ratio when anonymizing in-distribution images (and out-of-distribution images in parentheses) $\sim p_{\mathrm{test}}$}
\label{table:psnrs}
\tiny
\begin{tabular}{rrrrrr}
\toprule
FGSM  & PGD  & DDN & MISL & SpoC & Cama \\ \midrule
35.0 (35.0) & 34.6 (34.6) & 69.3 (88.1) & 35.2 (34.7) & 35.3 (34.8) & 35.7 (35.3)\\
\bottomrule
\end{tabular}
\end{table}
\setlength{\tabcolsep}{1.4pt}

\setlength{\tabcolsep}{4pt}
\begin{table}[t]
\centering
\caption{TSR (USR) in the \emph{same architecture complete overlap} setting when anonymizing (a) in-distribution images (cf. \Cref{table:id-resnet18}) and (b) out-of-distribution images (cf. \Cref{table:ood-resnet18})}
\subfloat[]{
\label{table:id-resnet18-add}
\tiny
\begin{tabular}{lrrrrrrr}
\toprule
& \multicolumn{6}{c}{{Preprocessor}} \\ \cmidrule(lr){2-7}
Attack & RGB & RGB+FD & FD & WW & CC & HP & Mean \\
\cmidrule(lr){1-1} \cmidrule(lr){2-7} \cmidrule(lr){8-8}
FGSM  & 15.3 (72.2) & 15.6 (75.3) & 12.9 (60.4) & 18.7 (73.1) & 16.6 (70.8) & 18.3 (75.8) & 16.2 (71.3)\\
PGD  & 17.9 (76.1) & 14.6 (69.1) & 16.0 (65.9) & 58.3 (88.7) & 17.0 (75.5) & 16.6 (67.5) & 23.4 (73.8)\\
DDN & 0.0 (0.2) & 0.0 (0.2) & 0.1 (0.2) & 0.0 (0.2) & 0.0 (0.2) & 0.1 (0.5) & 0.0 (0.2)\\ 
Cama  & \textbf{91.2} (\textbf{96.3}) & \textbf{86.8} (\textbf{87.7}) & \textbf{94.2} (\textbf{97.4}) & \textbf{97.3} (\textbf{98.0}) & \textbf{88.1} (\textbf{88.6}) & \textbf{89.7} (\textbf{92.6}) & \textbf{91.2} (\textbf{93.4})\\
\bottomrule
\end{tabular}
}

\centering
\subfloat[]{
\label{table:ood-resnet18-add}
\tiny
\begin{tabular}{lrrrrrrr}
\toprule
& \multicolumn{6}{c}{{Preprocessor}} \\ \cmidrule(lr){2-7}
Attack & RGB & RGB+FD & FD & WW & CC & HP & Mean \\
\cmidrule(lr){1-1} \cmidrule(lr){2-7} \cmidrule(lr){8-8}
FGSM  & 25.6 & 19.7 & 23.5 & 42.7 & 25.0 & 22.0 &  26.4\\
PGD  & 21.7 & 17.6 & 19.5 & 60.2 & 18.2 & 19.2 &  26.1\\
DDN & 16.8 & 17.4 & 18.3 & 19.2 & 18.3 & 18.3 &  18.0\\
Cama  & \textbf{98.5} & \textbf{97.7} & \textbf{94.2} & \textbf{98.1} & \textbf{98.8} & \textbf{97.8} & \textbf{97.5}\\
\bottomrule
\end{tabular}
}
\end{table}
\setlength{\tabcolsep}{1.4pt}

\setlength{\tabcolsep}{4pt}
\begin{table}[t]
\centering
\caption{TSR (USR) in the \emph{same architecture partial overlap} setting when anonymizing (a) in-distribution images (cf. \Cref{table:id-resnet18-exp}) and (b) out-of-distribution images (cf. \Cref{table:ood-resnet18-exp})}
\subfloat[]{
\label{table:id-resnet18-exp-add}
\tiny
\begin{tabular}{lrrrrrrr}
\toprule
& \multicolumn{6}{c}{{Preprocessor}} \\ \cmidrule(lr){2-7}
Attack & RGB & RGB+FD & FD & WW & CC & HP & Mean \\
\cmidrule(lr){1-1} \cmidrule(lr){2-7} \cmidrule(lr){8-8}
FGSM  & 10.7 (86.6) & 0.5 (94.8) & 3.2 (94.1) & 8.3 (86.0) & 0.6 (91.7) & 4.4 (95.5) &  4.6 (91.4) \\
PGD  & 13.8 (83.5) & 0.2 (96.7) & 1.1 (86.2) & 0.0 (100.0) & 0.2 (96.7) & 4.0 (74.0) &  3.2 (89.5)\\
DDN & 0.0 (0.2) & 0.0 (0.0) & 0.4 (0.8) & 0.0 (0.4) & 0.1 (0.8) & 0.0 (0.3) &  0.1 (0.4)
\\
Cama  & \textbf{81.8} (\textbf{91.3}) & \textbf{80.7} (\textbf{85.3}) & \textbf{94.9} (\textbf{98.3}) & \textbf{94.4} (\textbf{97.1}) & \textbf{82.0} (\textbf{93.5}) & \textbf{82.5} (\textbf{95.3}) & \textbf{86.0} (\textbf{93.5})\\
\bottomrule
\end{tabular}
}

\centering
\subfloat[]{
\label{table:ood-resnet18-exp-add}
\tiny
\begin{tabular}{lrrrrrrr}
\toprule
& \multicolumn{6}{c}{{Preprocessor}} \\ \cmidrule(lr){2-7}
Attack & RGB & RGB+FD & FD & WW & CC & HP & Mean \\
\cmidrule(lr){1-1} \cmidrule(lr){2-7} \cmidrule(lr){8-8}
FGSM  & 8.7 (88.1) & 0.1 (78.7) & 1.9 (78.7) & 6.9 (87.4) & 1.1 (71.1) & 8.5 (77.3) &  4.5 (80.2) \\
PGD  &  10.8 (85.6) & 0.1 (79.3) & 0.4 (66.7) & 0.0 (79.6) & 0.0 (75.4) & 4.5 (71.4) &  2.6 (76.3)\\
DDN & 0.2 (1.1) & 0.0 (0.0) & 0.0 (0.0) & 0.0 (0.0) & 0.0 (0.2) & 0.1 (0.8) &  0.1 (0.4) \\ 
Cama  & \textbf{86.6} (\textbf{95.0}) & \textbf{96.5} (\textbf{98.2}) & \textbf{95.5} (\textbf{98.8}) & \textbf{95.6} (\textbf{99.1}) & \textbf{83.6} (\textbf{95.5}) & \textbf{92.1} (\textbf{97.0}) & \textbf{91.7} (\textbf{97.3}) \\
\bottomrule
\end{tabular}
}
\end{table}
\setlength{\tabcolsep}{1.4pt}

\setlength{\tabcolsep}{4pt}
\begin{table}[t]
\centering
\caption{TSR (USR) in the \emph{architecture transfer complete overlap} setting when anonymizing (a) in-distribution images (cf. \Cref{table:id-others}) and (b) out-of-distribution images (cf. \Cref{table:ood-others})}
\subfloat[]{
\label{table:id-others-add}
\tiny
\begin{tabular}{lrrrrrrrr}
\toprule
& &\multicolumn{6}{c}{{Preprocessor}} \\ \cmidrule(lr){3-8}
 & Attack & RGB & RGB+FD & FD & WW & CC & HP & Mean \\
\cmidrule(lr){2-2} \cmidrule(lr){3-8} \cmidrule(lr){9-9}
\multirow{4}{*}{\rotatebox[origin=c]{90}{{R-50}}}
& FGSM  & 16.7 (82.7) & 16.6 (73.1) & 15.6 (74.2) & 18.5 (86.2) & 17.0 (78.1) & 18.2 (82.3) &  17.1 (79.4) \\
& PGD  & 14.7 (67.1) & 20.5 (74.5) & 16.3 (81.6) & 36.0 (75.0) & 16.6 (81.6) & 16.0 (80.2) &  20.0 (76.7)\\
& DDN & 0.1 (0.3) & 0.0 (0.0) & 0.0 (0.2) & 0.0 (0.0) & 0.0 (0.2) & 0.0 (0.2) &  0.0 (0.2) \\
& Cama  & \textbf{87.9} (\textbf{92.3}) & \textbf{79.8} (\textbf{83.7}) & \textbf{92.0} (\textbf{92.5}) & \textbf{95.1} (\textbf{95.9}) & \textbf{93.9} (\textbf{95.9}) & \textbf{89.3} (\textbf{89.9}) & \textbf{89.7} (\textbf{91.7}) \\

\midrule
\multirow{4}{*}{\rotatebox[origin=c]{90}{{D-100}}}
& FGSM  & 17.5 (83.4) & 16.6 (71.0) & 16.0 (79.8) & 18.0 (80.1) & 16.4 (76.4) & 19.0 (81.4) &  17.2 (78.7) \\
& PGD  & 26.4 (59.2) & 19.4 (71.1) & 16.2 (79.4) & 24.2 (77.9) & 16.2 (79.7) & 26.1 (74.6) &  21.4 (73.6) \\
& DDN & 0.0 (0.0) & 0.0 (0.0) & 0.0 (0.2) & 0.0 (0.2) & 0.0 (0.2) & 0.1 (0.3) &  0.0 (0.2) \\
& Cama  & \textbf{97.8} (\textbf{98.7}) & \textbf{97.4} (\textbf{97.8}) & \textbf{98.8} (\textbf{99.3}) & \textbf{99.5} (\textbf{99.8}) & \textbf{98.6} (98.9) & \textbf{96.9} (\textbf{99.5}) & \textbf{98.2} (\textbf{99.0}) \\

\midrule
\multirow{4}{*}{\rotatebox[origin=c]{90}{{V-16}}}

& FGSM  & 19.8 (93.3) & 16.2 (78.3) & 16.0 (80.9) & 16.7 (83.3) & 16.7 (79.5) & 21.4 (80.6) &  17.8 (82.6)\\
& PGD  & 23.1 (77.7) & 15.3 (77.1) & 16.3 (81.1) & 17.1 (83.4) & 16.4 (80.3) & 34.1 (76.3) &  20.4 (79.3)\\
& DDN & 0.4 (1.7) & 0.1 (0.2) & 0.0 (0.2) & 0.3 (0.9) & 0.0 (0.2) & 0.0 (0.2) &  0.1 (0.6) \\
& Cama  & \textbf{88.0} (\textbf{95.9}) & \textbf{99.0} (\textbf{99.3}) & \textbf{98.1} (\textbf{98.8}) & \textbf{87.4} (\textbf{96.6}) & \textbf{98.7} (\textbf{99.3}) & \textbf{98.4} (\textbf{98.8}) & \textbf{94.9} (\textbf{98.1}) \\
\bottomrule
\end{tabular}
}

\centering
\subfloat[]{
\label{table:ood-others-add}
\tiny
\begin{tabular}{lrrrrrrrr}
\toprule
& &\multicolumn{6}{c}{{Preprocessor}} \\ \cmidrule(lr){3-8}
 & Attack & RGB & RGB+FD & FD & WW & CC & HP & Mean \\
\cmidrule(lr){2-2} \cmidrule(lr){3-8} \cmidrule(lr){9-9}
\multirow{4}{*}{\rotatebox[origin=c]{90}{{R-50}}}
& FGSM  & 20.3 & 25.7 & 17.8 & 22.1 & 18.2 & 19.5 &  20.6\\
& PGD  & 19.0 & 21.3 & 17.0 & 38.2 & 16.8 & 17.4 &  21.6\\
& DDN & 16.9 & 18.4 & 18.2 & 18.1 & 18.3 & 18.6 &  18.1\\
& Cama  & \textbf{95.9} & \textbf{96.7} & \textbf{99.2} & \textbf{99.4} & \textbf{98.9} & \textbf{99.1} & \textbf{98.2}\\
\midrule
\multirow{4}{*}{\rotatebox[origin=c]{90}{{D-100}}}
& FGSM  &19.4 & 26.1 & 18.7 & 21.9 & 19.9 & 31.7 &  23.0 \\
& PGD  & 28.6 & 20.4 & 17.2 & 24.1 & 17.2 & 31.6 &  23.2\\
& DDN & 16.9 & 18.3 & 19.4 & 19.4 & 17.4 & 18.2 &  18.3\\
& Cama  & \textbf{99.1} & \textbf{99.6} & \textbf{99.6} & \textbf{99.7} & \textbf{99.3} & \textbf{98.6} & \textbf{99.3} \\
\midrule
\multirow{4}{*}{\rotatebox[origin=c]{90}{{V-16}}}

& FGSM  & 19.6 & 19.0 & 17.6 & 16.8 & 19.1 & 35.1 &  21.2\\
& PGD  & 28.4 & 16.9 & 16.9 & 17.0 & 17.0 & 36.6 &  22.1\\
& DDN & 17.4 & 17.2 & 17.2 & 17.6 & 19.3 & 17.7 &  17.7\\
& Cama  & \textbf{91.6} & \textbf{99.8} & \textbf{98.2} & \textbf{91.7} & \textbf{99.3} & \textbf{99.3} & \textbf{96.6}  \\
\bottomrule
\end{tabular}
}
\end{table}
\setlength{\tabcolsep}{1.4pt}

\setlength{\tabcolsep}{4pt}
\begin{table}[t]
\centering
\caption{TSR (USR) in the \emph{architecture transfer partial overlap} setting when anonymizing (a) in-distribution images (cf. \Cref{table:id-others-exp}) and (b) out-of-distribution images (cf. \Cref{table:ood-others-exp})}
\subfloat[]{
\label{table:id-others-exp-add}
\tiny
\begin{tabular}{lrrrrrrrr}
\toprule
& &\multicolumn{6}{c}{{Preprocessor}} \\ \cmidrule(lr){3-8}
 & Attack & RGB & RGB+FD & FD & WW & CC & HP & Mean \\
\cmidrule(lr){2-2} \cmidrule(lr){3-8} \cmidrule(lr){9-9}
\multirow{4}{*}{\rotatebox[origin=c]{90}{{R-50}}}
& FGSM  & 0.5 (96.9) & 8.5 (86.2) & 0.2 (97.6) & 11.5 (85.2) & 12.8 (86.0) & 1.0 (95.4) &  5.8 (91.2)\\
& PGD  & 0.2 (96.3) & 5.6 (87.9) & 0.0 (99.4) & 16.1 (77.4) & 6.6 (93.7) & 0.5 (95.7) &  4.8 (91.7)\\
& DDN  & 0.0 (0.3) & 0.1 (6.5) & 0.0 (0.2) & 0.0 (0.2) & 0.4 (4.5) & 0.0 (0.7) &  0.1 (2.1) \\
& Cama  & \textbf{92.6} (\textbf{97.6}) & \textbf{73.3} (\textbf{88.2}) & \textbf{83.3} (\textbf{98.2}) & \textbf{91.9} (\textbf{99.3}) & \textbf{81.9} (\textbf{96.4}) & \textbf{75.3} (\textbf{88.4}) & \textbf{83.0} (\textbf{94.7}) \\
\midrule
\multirow{4}{*}{\rotatebox[origin=c]{90}{{D-100}}}
& FGSM  & 0.4 (98.3) & 2.5 (96.5) & 0.1 (99.5) & 1.7 (100.0) & 8.5 (86.5) & 10.9 (81.5) &  4.0 (93.7)\\
& PGD  & 2.3 (90.0) & 2.7 (89.6) & 0.0 (98.2) & 0.3 (99.2) & 0.8 (96.0) & 16.1 (84.2) &  3.7 (92.9) \\
& DDN  & 0.0 (0.0) & 0.0 (0.0) & 0.0 (0.1) & 0.0 (0.0) & 0.0 (0.2) & 0.0 (0.2) &  0.0 (0.1)\\
& Cama  & \textbf{88.3} (\textbf{95.0}) & \textbf{96.7} (\textbf{98.6}) & \textbf{96.3} (\textbf{99.5}) & \textbf{96.9} (\textbf{99.2}) & \textbf{88.9} (\textbf{97.4}) & \textbf{88.9} (\textbf{99.2}) & \textbf{92.7} (\textbf{98.2}) \\
\midrule
\multirow{4}{*}{\rotatebox[origin=c]{90}{{V-16}}}
& FGSM  & 0.0 (99.8) & 4.2 (87.4) & 16.4 (81.9) & 7.0 (96.4) & 0.3 (97.9) & 7.0 (85.8) &  5.8 (91.5)\\
& PGD  &
0.0 (99.2) & 1.0 (97.9) & 15.9 (78.8) & 0.8 (96.2) & 0.2 (97.8) & 5.2 (79.1) &  3.8 (91.5) \\
& DDN  & 0.0 (0.0) & 0.0 (0.0) & 0.0 (0.1) & 0.0 (0.1) & 0.0 (0.2) & 0.0 (0.2) &  0.0 (0.1)\\
& Cama  & \textbf{98.3} (\textbf{99.5}) & \textbf{95.1} (\textbf{98.4}) & \textbf{94.5} (\textbf{99.0}) & \textbf{94.3} (\textbf{99.4}) & \textbf{97.3} (\textbf{98.9}) & \textbf{92.9} (\textbf{99.2}) & \textbf{95.4} (\textbf{99.1})\\
\bottomrule
\end{tabular}
}

\centering
\subfloat[]{
\label{table:ood-others-exp-add}
\tiny
\begin{tabular}{lrrrrrrrr}
\toprule
& &\multicolumn{6}{c}{{Preprocessor}} \\ \cmidrule(lr){3-8}
 & Attack & RGB & RGB+FD & FD & WW & CC & HP & Mean \\
\cmidrule(lr){2-2} \cmidrule(lr){3-8} \cmidrule(lr){9-9}
\multirow{3}{*}{\rotatebox[origin=c]{90}{{R-50}}}
& FGSM  & 1.0 (76.4) & 11.0 (92.5) & 0.0 (78.6) & 16.4 (91.9) & 15.1 (95.5) & 1.2 (76.1) &  7.4 (85.2) \\
& PGD  & 0.3 (61.3) & 7.8 (81.6) & 0.0 (78.9) & 16.8 (92.0) & 4.7 (86.6) & 0.0 (74.2) &  4.9 (79.1)\\
& DDN  & 0.3 (1.8) & 0.1 (1.5) & 0.0 (0.2) & 0.0 (0.0) & 1.6 (15.6) & 0.1 (0.4) &  0.4 (3.2)\\
& Cama  & \textbf{96.1} (\textbf{97.9}) & \textbf{88.4} (\textbf{94.4}) & \textbf{88.7} (\textbf{96.0}) & \textbf{97.5} (\textbf{99.1}) & \textbf{81.3} (\textbf{93.6}) & \textbf{86.8} (\textbf{97.1}) & \textbf{89.8} (\textbf{96.4})\\
\midrule
\multirow{3}{*}{\rotatebox[origin=c]{90}{{D-100}}}
& FGSM  & 0.5 (77.1) & 1.6 (83.1) & 0.7 (78.5) & 0.3 (80.6) & 7.9 (80.6) & 16.5 (85.4) &  4.6 (80.9)\\
& PGD  & 1.2 (66.9) & 2.4 (80.8) & 0.0 (73.2) & 0.1 (74.6) & 0.2 (78.6) & 16.0 (84.0) &  3.3 (76.4)\\
& DDN  & 0.0 (0.0) & 0.0 (0.0) & 0.0 (0.0) & 0.0 (0.0) & 0.0 (0.0) & 0.0 (0.2) &  0.0 (0.0)\\
& Cama  & \textbf{92.3} (\textbf{95.5}) & \textbf{97.8} (\textbf{98.5}) & \textbf{95.2} (\textbf{98.3}) & \textbf{97.3} (\textbf{98.8}) & \textbf{91.2} (\textbf{95.6}) & \textbf{94.7} (\textbf{98.0}) & \textbf{94.8} (\textbf{97.4})\\
\midrule
\multirow{3}{*}{\rotatebox[origin=c]{90}{{V-16}}}
& FGSM  & 0.1 (81.8) & 5.6 (82.6) & 17.1 (98.1) & 2.5 (80.4) & 0.1 (77.1) & 11.1 (74.0) &  6.1 (82.3)\\
& PGD  & 0.1 (79.3) & 0.6 (81.6) & 16.8 (99.4) & 0.1 (79.2) & 0.1 (78.8) & 4.6 (66.9) &  3.7 (80.9)\\
& DDN  & 0.0 (0.0) & 0.0 (0.0) & 0.0 (0.0) & 0.0 (0.0) & 0.0 (0.2) & 0.0 (0.0) &  0.0 (0.0)\\
& Cama  & \textbf{98.9} (\textbf{99.6}) & \textbf{97.6} (\textbf{98.8}) & \textbf{93.0} (\textbf{97.4}) & \textbf{95.4} (\textbf{97.5}) & \textbf{96.4} (\textbf{98.8}) & \textbf{95.9} (\textbf{97.9}) & \textbf{96.2} (\textbf{98.3})\\
\bottomrule
\end{tabular}
}
\end{table}
\setlength{\tabcolsep}{1.4pt}

\end{document}